\documentclass[12pt]{article}
\usepackage{aaspp4}
\usepackage[final]{graphicx}

\graphicspath{{fig/}}
\begin{document}

\newcommand{\ttt}[1]{\ensuremath{\times 10^{#1}}}
\newcommand{\alp}{\ensuremath{\alpha}}
\newcommand{\tnine}{\ensuremath{{T}_{9}}}
\newcommand{\tninei}{\ensuremath{{T}_{9i}}}
\newcommand{\Abar}{\ensuremath{\bar A}}
\newcommand{\nuc}[2]{\ensuremath{\mathrm {^{#2}#1}}}
\newcommand{\gcc}{\ensuremath{\mathrm{\thinspace g \thinspace cm^{-3}}}}
\newcommand{\kel}{\ensuremath{\mathrm{\thinspace K}}}
\newcommand{\etal}{ et~al.}
\newcommand{\canjphys}{Can. J. Phys.}
\newcommand{\revmodphys}{Rev. Mod. Phys.}

\title{Silicon Burning II: Quasi-Equilibrium and Explosive Burning}

\author{W. Raphael Hix\altaffilmark{1,2,3,4}
\& Friedrich-Karl Thielemann\altaffilmark{3,5}}
\altaffiltext{1}{Joint Institute for Heavy Ion Research,  Oak Ridge
National Laboratory, 
Oak Ridge, TN 37831-6374}
\altaffiltext{2}{Department of Physics and Astronomy, University of
Tennessee, Knoxville, TN 37996-1200}
\altaffiltext{3}{Physics Division, Oak Ridge National Laboratory, 
Oak Ridge, TN 37831-6373}
\altaffiltext{4}{Department of Astronomy, University of Texas, Austin TX, 
78712}
\altaffiltext{5}{Department of Physics and Astronomy, University of Basel, 
Klingelberstrasse 82, CH-4056 Basel Switzerland}

\slugcomment{ \tiny The submitted manuscript has been authored by a
contractor of the U.S.\ Government under contract
DE-AC05-96OR22464.  Accordingly, the U.S.\ Government
retains a nonexclusive, royalty-free license to
publish or reproduce the published form of this
contribution, or allow
others to do so, for U.S.\ Government purposes.}

\begin{abstract}
Having examined the application of quasi-equilibrium to hydrostatic silicon 
burning in Paper I of this series, \cite{HiTh96}, we now turn our attention 
to explosive silicon burning.  Previous authors have shown that for 
material which is heated to high temperature by a passing shock and then 
cooled by adiabatic expansion, the results can be divided into three broad 
categories; \emph{incomplete burning}, \emph{normal freezeout} and 
\emph{\alp-rich freezeout}, with the outcome depending on the temperature, 
density and cooling timescale.  In all three cases, we find that the 
important abundances obey quasi-equilibrium for temperatures greater than 
approximately $3 \ttt{9} \kel$, with relatively little nucleosynthesis 
occurring following the breakdown of quasi-equilibrium.  We will show that 
quasi-equilibrium provides better abundance estimates than global nuclear 
statistical equilibrium, even for normal freezeout and particularly for 
\alp-rich freezeout.  We will also examine the accuracy with which the 
final nuclear abundances can be estimated from quasi-equilibrium.
\end{abstract}


\section{Introduction}

Because the products of hydrostatic silicon burning are trapped deep in the 
potential well of their parent star, it is only by explosion that the 
interstellar medium is enriched in intermediate mass and iron peak 
elements.  In core collapse supernovae, the neutrino driven shock passes 
through the silicon and oxygen shells, heating the material and driving at 
least a portion outward.  For thermonuclear supernovae, the degenerate 
ignition of carbon and oxygen provides both the fuel for silicon burning 
and the energy required to unbind this material from its parent white 
dwarf.  In both cases, matter is heated rapidly to temperatures sufficient 
to destroy silicon and expelled outward, to expand and cool adiabatically.  
Thus, understanding the chemical evolution of the intermediate mass and 
iron peak elements, and how the formation of these elements affects the 
supernovae which produce them, requires understanding the behavior of 
silicon burning under such explosive conditions.  Recent calculations of 
both core collapse and thermonuclear supernovae seem to indicate that, in 
both cases, silicon burning occurs in material subject to strong 
hydrodynamic instabilities, prompting two- and ultimately three-dimensional 
simulations.  See, for example, \cite{BaAr94} for discussion of convection 
prior to core collapse; \cite{HBHFC94}, \cite{BuHF95}, \cite{JaMu95}, and 
\cite{Meea98} for discussion of core collapse instabilities; and 
\cite{Khok93,Khok95}, and \cite{NiHi95} for discussion of turbulent 
flames in thermonuclear supernovae.  While it has been possible to perform 
simulations of silicon burning using large nuclear networks within 
spherically symmetric models comprised of hundreds of zones, the need to 
understand silicon burning in these multi-dimensional contexts, with the 
attendant squaring and cubing of the number of zones, demands more 
efficient calculation.  This provides an additional motivation for our 
study, as we seek to leverage our understanding of silicon burning to 
reduce its computational cost.

In a previous paper, \cite{HiTh96}, henceforth HT96, we described how the 
physics of silicon burning is dominated by partial equilibria among groups 
of nuclei well before complete nuclear statistical equilibrium (NSE) is 
established.  \cite{BCF68} introduced the concept of quasi-equilibrium 
(QSE), contending that the solar system abundances of nuclei between neon 
and the iron peak could be explained by a single quasi-equilibrium group 
which stretched from magnesium through the iron peak but failed to reach 
NSE. This mechanism unified the \alp- and e-processes of \cite{B2FH57} and 
reconciled moderately well with the nuclear network calculations of 
\cite{TrCG66}.  QSE is a local equilibrium with respect to strong and 
electromagnetic reactions, i.e. the exchange of neutrons, protons, 
\alp-particles and photons.  In HT96, we demonstrated that the abundance of 
a nucleus in QSE with \nuc{Si}{28}, for example, is given by
\begin{equation}
Y_{QSE}(^AZ) = \left(C(^AZ) \over C({\rm ^{28}Si}) 
\right) Y({\rm ^{28}Si}) {Y_n}^{N-14} {Y_p}^{Z-14} \ , \label{eq:qse}
\end{equation}
where $Y_n$, $Y_p$, and $Y({\rm ^{28}Si})$ are the abundances of free 
neutrons, free protons, and \nuc{Si}{28}.  The thermodynamic conditions and 
nuclear properties are contained within the expressions
\begin{equation}
C(^AZ)= {G(^AZ) \over 2^A} {\left(\rho N_A \over \theta \right)}^{A-1} 
A^{3 \over 2} \exp {\left( B(^AZ) \over {k_B T} \right)} \label{eq:cnse}
\end{equation}
which has been defined for later convenience, with
\begin{equation}
\theta = \left( {m_u k_B T \over 2 \pi \hbar^2 } \right)^{3/2} \  .
\end{equation}
$G(^AZ)$ and $B(^AZ)$ are the partition function and binding energy of the 
nucleus $^AZ$, $N_A$ is Avagadro's number, $k_B$ is Boltzmann's constant, 
and $\rho$ and $T$ are the density and temperature of the plasma.  It is 
worth noting that, due to the $\exp (B(^AZ)/{k_B T})$ term, declining 
temperature favors the more tightly bound nuclei.  Furthermore, for the 
thermodynamic conditions common to silicon burning, the nuclear binding 
energy, $B(^AZ)$, must be corrected for the effects of Coulomb screening 
(\cite{HTFT99}).

As Eq.~\ref{eq:qse} demonstrates, the abundance of any nucleus in 
quasi-equilibrium can be expressed as a function of just three abundances, 
the thermodynamic conditions ($\rho$, \tnine) and the properties of that 
nucleus, subsumed here within $C(^AZ)$.  Thus the evolution of the 
abundances of \nuc{Si}{28} and the free nucleons defines the behavior of 
all the nuclei in QSE with \nuc{Si}{28}.  One can naturally construct 
similar equations relative to nuclei other than \nuc{Si}{28}, though for 
all nuclei within the silicon QSE group, all such equations reduce to 
Eq.~\ref{eq:qse}.  This expression for quasi-equilibrium, Eq.~\ref{eq:qse}, 
is also identical to those defined by \cite{BCF68}, \cite{WAC73} and
\cite{MeKC96}, provided the \alp-particles, protons and neutrons are 
internally in equilibrium.  We find that, by the time quasi-equilibrium 
is established in the vicinity of silicon, these light nuclei consistently 
have formed their own QSE group, with abundances in this light QSE group 
obeying the NSE relation (Eq.\ref{eq:nse}).  These light nuclei remain 
in QSE until the bulk of the photodisintegration rates freezeout for 
$\tnine \approx 3$.  

In contrast to quasi-equilibrium, nuclear statistical equilibrium is a 
global equilibrium.  From arguments based on chemical potentials or 
detailed balance of reactions (see, for example, \cite{HTFT99}, 
\cite{HWEE85}, \cite{Clay83}, or \cite{ClTa65}), an expression for the 
equilibrium abundances of all nuclei can be derived,
\begin{equation}
Y_{NSE}(^AZ) = C(^AZ){Y_n}^{N} {Y_p}^{Z} \ . \label{eq:nse}
\end{equation}
Here $C(^AZ)$ is identical to Eq.~\ref{eq:cnse}, and $Y_n$ and $Y_p$ are 
the free neutron and free proton abundances.  Thus in NSE, the abundances 
of all nuclei are functions of only two abundances.  Substitution of 
Eq.~\ref{eq:nse}, evaluated for \nuc{Si}{28}, into Eq.~\ref{eq:qse}, 
reveals that for all nuclei Eq.~\ref{eq:qse} reduces to Eq.~\ref{eq:nse} if 
\nuc{Si}{28} has reached NSE. Thus failure of a quasi-equilibrium group to 
reach NSE implies that while the members of the group have achieved 
equilibrium with respect to the exchange of free nucleons and 
\alp-particles, they have not achieved global equilibrium with the free 
nucleons.

In HT96, in order to separate the details of silicon burning from the 
vagaries of hydrodynamics, we considered only constant thermodynamic 
conditions.  Further, we limited ourselves to conditions appropriate for 
hydrostatic silicon burning in the late stages of stellar evolution.  While 
the thermodynamic variations within the core of a star undergoing 
hydrostatic silicon burning occur slowly, this is certainly not true for 
convective or explosive silicon burning.  If quasi-equilibrium is to be 
useful in understanding and simplifying the modeling of silicon burning 
within a hydrodynamic supernova model, it is necessary to demonstrate that 
its applicability is not limited to constant conditions.  As a 
representative analytic model for silicon burning occurring as a result of 
shock heating, we will consider a mass zone which is instantaneously heated 
by a passing shock to some peak initial temperature, $\tninei$, and 
density, $\rho_i$, and then expands and cools adiabatically.  Following the 
approximation introduced by \cite{FoHo64}, the expansion timescale (equal 
to the free fall timescale) is
\begin{equation}
\tau_{\rm HD} = \left(24 \pi G \rho \right)^{-1/2} = 446 \rho_6^{-1/2} 
\mathrm {ms}, 
\end{equation}
with the time dependence of the density and temperature of this
radiation dominated gas given by
\begin{eqnarray}
\rho(t) &= &\rho_i\exp \left( -t / \tau_{\mathrm {HD}} \right), \nonumber \\
T_9(t) &= &T_{9i} \exp \left( -t / 3 \tau_{\mathrm {HD}} \right), 
\end{eqnarray}
assuming the adiabatic exponent $\gamma= 4/3$.  This model provides a 
strong but simple test of the behavior of quasi-equilibrium under varying 
conditions.  

\begin{figure}[t]
    \centering
    \includegraphics[angle=-90,width=\textwidth]{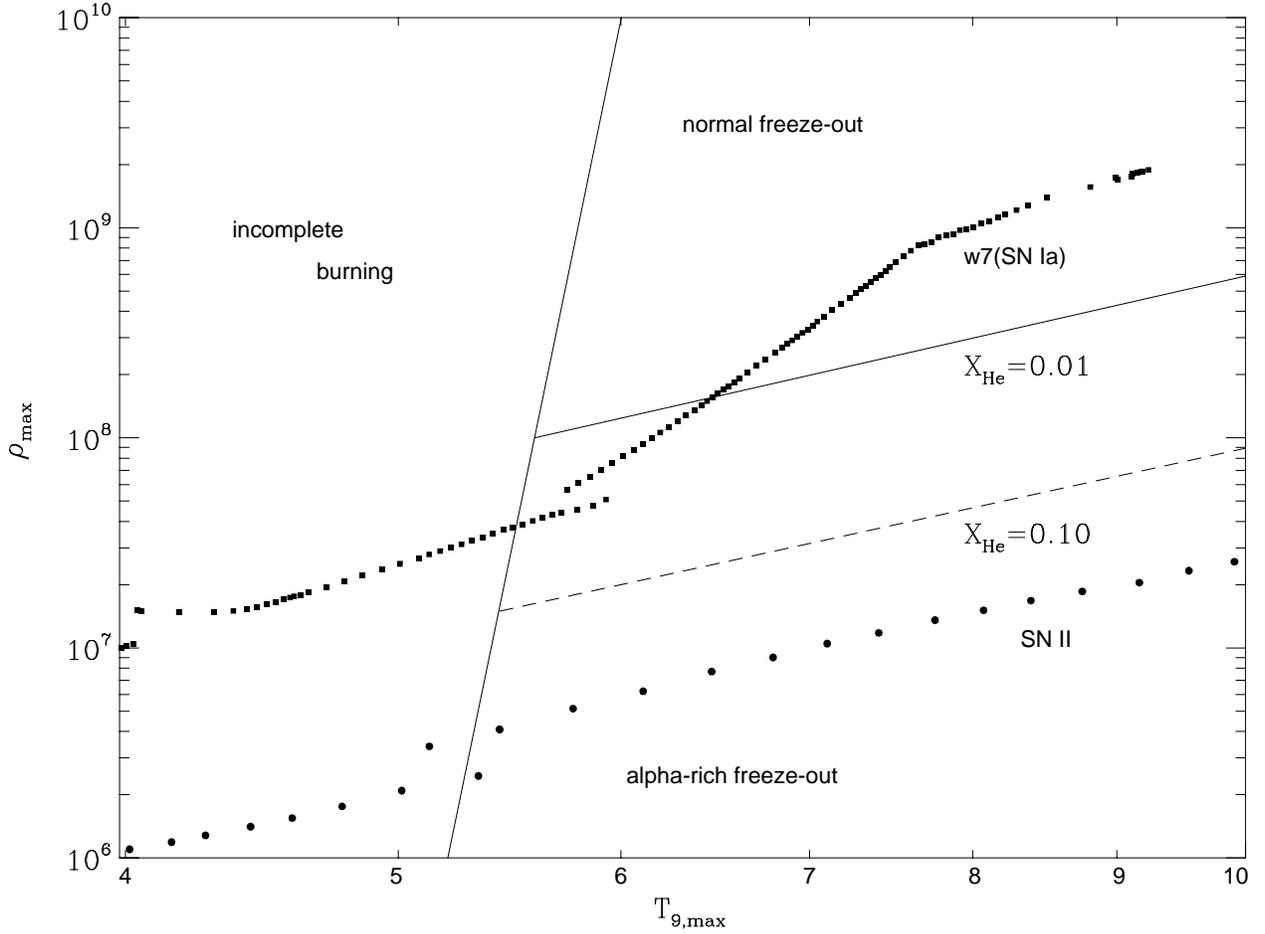}
    \caption{Outcome of explosive silicon burning as a function
    of peak temperature and density.  Contour lines of constant $^4$He mass
    fractions in complete burning are given for levels of 1 and 10\%. They 
    coincide with lines of constant radiation entropy per gram of matter. 
    For comparision also the maximum $\rho-T$-conditions of individual mass 
    zones in type Ia and type II supernovae are indicated.}
    \label{fig:freeze_cond}
\end{figure}

Previous authors, beginning with \cite{WAC73}, have shown that for 
explosive burning, the $\tninei, \rho_i$ plane is divided into 3 regions.  
Figure 1 portrays this division for material with $Y_e \lesssim .5$, while 
also showing the remnant helium mass fraction.  The squares plotted in this 
figure indicate the range of conditions found in the W7 model for Type Ia 
supernovae (see \cite{ThNY86}), while the circles show conditions typical 
for Type II supernovae models (see, for example, \cite{WoWe95} and 
\cite{ThNH96}).  In the first region, roughly defined by $\tninei<5$, 
cooling due to the expansion halts the destruction of silicon prematurely, 
resulting in a larger remnant concentration of intermediate mass elements 
than nuclear statistical equilibrium would suggest.  Quite naturally, this 
is referred to as \emph{incomplete silicon burning}.  For higher 
temperatures silicon is exhausted, and NSE is reached before the reactions 
freeze out.  For moderately large densities ($\rho_i \gtrsim 10^{8} \gcc$), 
the abundances of the light QSE group nuclei are relatively small, allowing 
these light nuclei to pass through the A=5-8 bottleneck and form heavier 
nuclei as the temperature, and therefore their equilibrium population, 
drops.  When this occurs, NSE persists up until freezeout at $\tnine \sim 
3$, resulting in a \emph{normal freezeout}.  At lower densities the light 
group nuclei, especially \nuc{He}{4}, contain a large portion of the 
nuclear mass fraction.  This larger NSE abundance of light nuclei, coupled 
with the reduced reaction flow through the bottleneck due to the quadratic 
density dependence of the rates for the $\alp+\alp+\alp\rightarrow 
\nuc{C}{12}$ and $\alp+\alp+n\rightarrow\nuc{Be}{9}$ reactions which bridge 
this mass gap, prevents the incorporation of all of these light nuclei into 
heavier nuclei on the timescale of the cooling.  Since the same binding 
energy considerations which favor the iron peak nuclei in NSE also favor 
\alp-particles over free nucleons, the light group mass is dominated by 
\alp-particles, hence this is termed \emph{\alp-rich freezeout}.  This 
paper will analyze and demonstrate the applicability of quasi-equilibrium 
for each of the 3 results of explosive silicon burning detailed above.

\section{Incomplete Burning}

We begin with incomplete silicon burning and an example where $\tninei=5$, 
$\rho= 10^{9} \gcc$ and $Y_e=.48$.  Figure 2a displays the ratio of the 
network abundances to their silicon QSE abundances plotted as a function of 
atomic mass after an elapsed time of $9.6 \times 10^{-7}$ seconds.  This 
figure is reminiscent of the figures discussed in HT96.  As in HT96,
these QSE abundances are calculated from the network values of $Y_n$,
$Y_p$, and $Y(\nuc{Si}{28})$, not on an independant QSE-based
calculation.  In Figures 2, 3 \& 
5, the filled triangles, squares and circles mark the core nuclei of the 3 
QSE group discussed in that paper.  These same 3 QSE groups are clearly 
present here, one composed of the light nuclei, the second centered on 
silicon and the third containing the iron peak nuclei.  One point to note 
is the apparent displacement of \nuc{He}{3} from the light group.  We 
believe that this is due to a limitation of our reaction set, which 
contains many fewer reactions involving \nuc{He}{3} than reactions 
involving \nuc{He}{4}.  If the corresponding reactions for \nuc{He}{3}, 
which are important only to the abundance of \nuc{He}{3}, were included, 
\nuc{He}{3} would also be in QSE with the free nucleons.  Clearly 
quasi-equilibrium is well established early in this calculation.  We then 
must ask, Under what conditions does quasi-equilibrium fail to be a good 
approximation to the network abundances?  Figure 2b is similar to Figure 
2a, with an elapsed time of $4.4 \times 10^{-3}$ seconds.  By this time, 
\tnine\ has dropped to 4.5 and $\rho= 7.3 \times 10^{8} \gcc$.  The 
evolution of the QSE groups follows the pattern discussed under constant 
thermodynamic conditions in HT96, with both the light and iron peak groups 
converging toward the silicon group.  In particular, the convergence of the 
light and silicon QSE groups reflects the approach of the network 
abundances to those calculated from NSE for the current thermodynamic 
conditions.

\begin{figure}[t]
    \centering
    \includegraphics[angle=90,width=\textwidth]{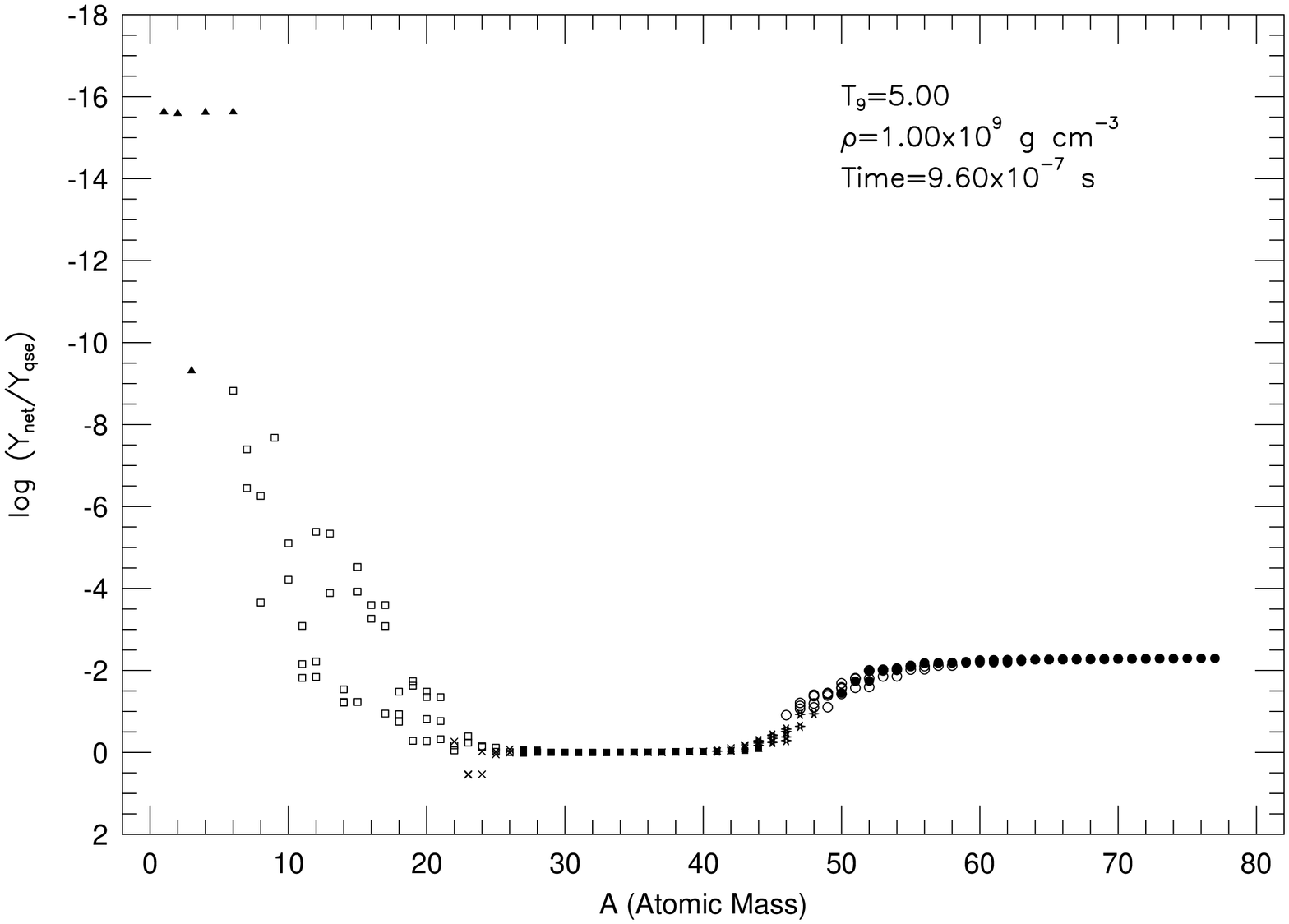}
    \caption{(a) Examination of the QSE group structure for $\tninei=5$, 
	$\rho_i= 10^9 \gcc$, and $Y_e=.48$ after elapsed time of $9.6 
	\times 10^{-7}$ seconds.  $Y_{net}$ are the network abundances, while 
	$Y_{qse}$ are abundances calculated assuming QSE from the network 
	abundances of n, p, \nuc{Si}{28}.}
    \label{fig:2a}
    \addtocounter{figure}{-1}
\end{figure}

\begin{figure}[ht]
    \centering
    \includegraphics[angle=90,width=\textwidth]{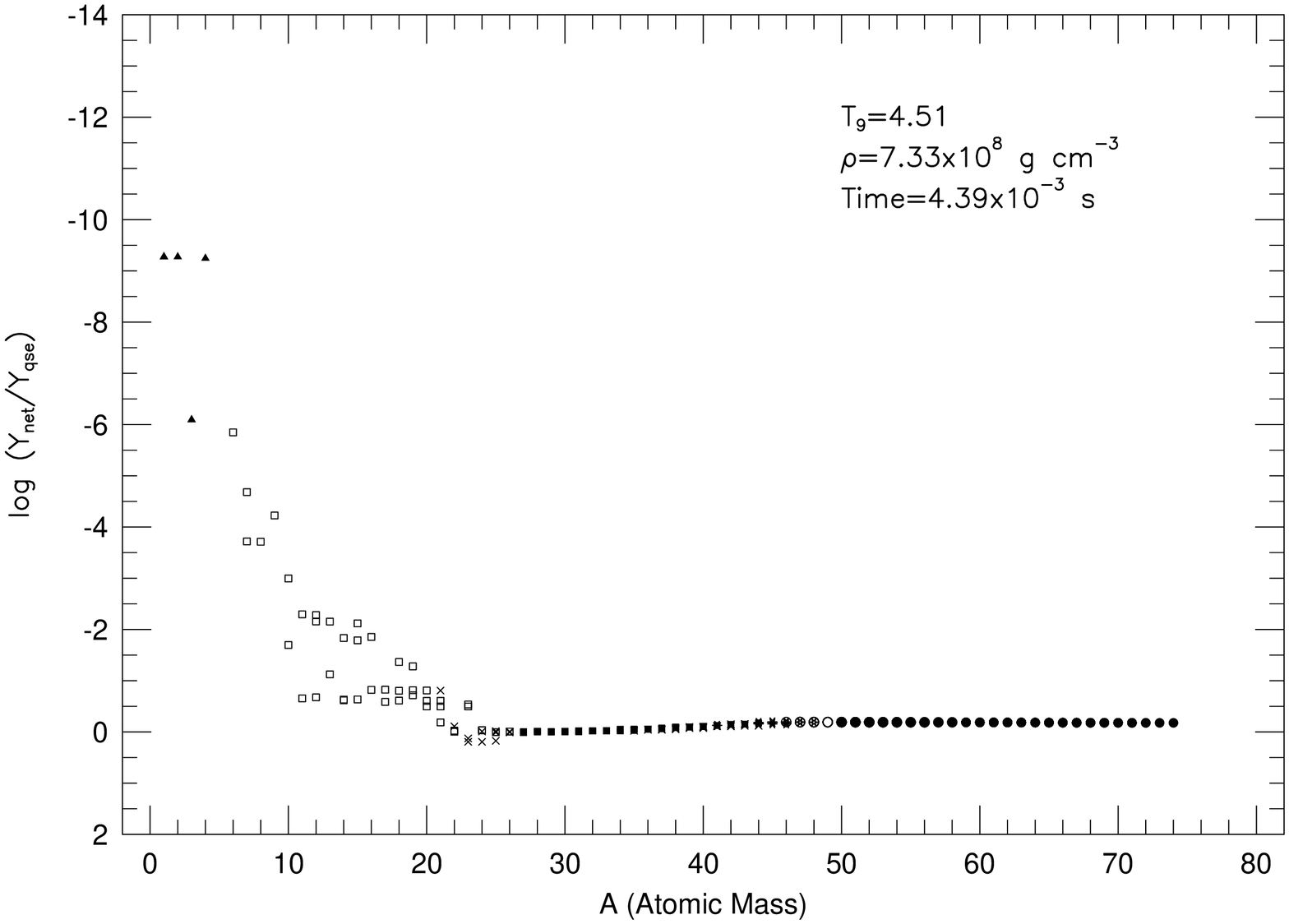}
    \caption{(b) Examination of the QSE group structure for $\tninei=5$, 
        $\rho_i= 10^9 \gcc$, and $Y_e=.48$ after elapsed time of $4.4 
        \times 10^{-3}$ ($\tnine=4.5$, $\rho=7.3 \times 10^{8} \gcc$).  
        $Y_{net}$ are the network abundances, while $Y_{qse}$ are abundances 
        calculated assuming QSE from the network abundances of n, p, 
        \nuc{Si}{28}.}
    \label{fig:2b}
    \addtocounter{figure}{-1}
\end{figure}

However, as time continues to elapse, this convergence reverses with the 
network abundances diverging from the then current NSE abundances.  As we 
discussed in \S1, in equilibrium, decreasing temperature favors the more 
bound members of a group of nuclei.  In this case, decreasing temperature 
raises the quasi-equilibrium abundances of the iron peak nuclei relative to 
all other nuclei.  Over this span of time, though the abundances of the 
iron peak nuclei are increasing, their QSE abundances are increasing more 
rapidly, moving the iron peak QSE group away from equilibrium with silicon.  
Similarly, because of the shift in favor of heavier nuclei, though the 
network abundance of silicon is declining, the equilibrium abundance of 
silicon is declining more rapidly, leaving silicon further from equilibrium 
with the free nucleons.  In spite of these movements, quasi-equilibrium is 
still a good approximation, with the abundances in the iron peak group, 
which comprises 80\% of the mass by the time $9.5 \times 10^{-3}$ seconds 
have passed (corresponding to $\tnine=4$ and $\rho=5.1 \times 10^{8} 
\gcc$), having a spread of a few \% from quasi-equilibrium, while those in 
the silicon group (comprising 10\% of the mass) have a spread of roughly 
15\%.

\begin{figure}[ht]
    \centering
    \includegraphics[angle=90,width=\textwidth]{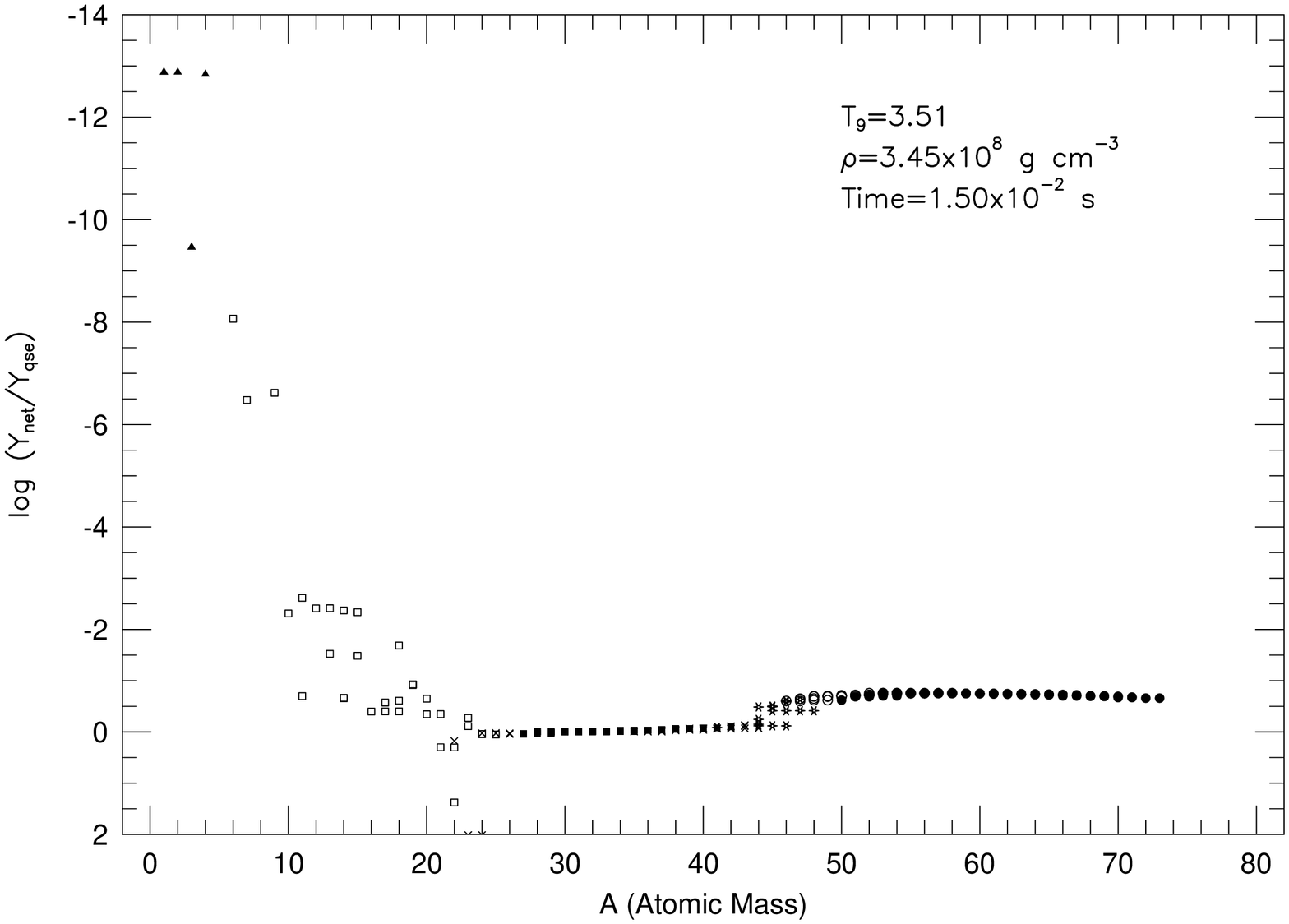}
    \caption{(c) Examination of the QSE group structure for $\tninei=5$, 
	$\rho_i= 10^9 \gcc$, and $Y_e=.48$ after elapsed time of $1.5 
	\times 10^{-2}$ seconds ($\tnine=3.5$, $\rho=3.5 \times 10^{8} \gcc$).  
	$Y_{net}$ are the network abundances, while $Y_{qse}$ are abundances 
	calculated assuming QSE from the network abundances of n, p, 
	\nuc{Si}{28}.}
    \label{fig:2c}
    \addtocounter{figure}{-1}
\end{figure}

\begin{figure}[ht]
    \centering
    \includegraphics[angle=90,width=\textwidth]{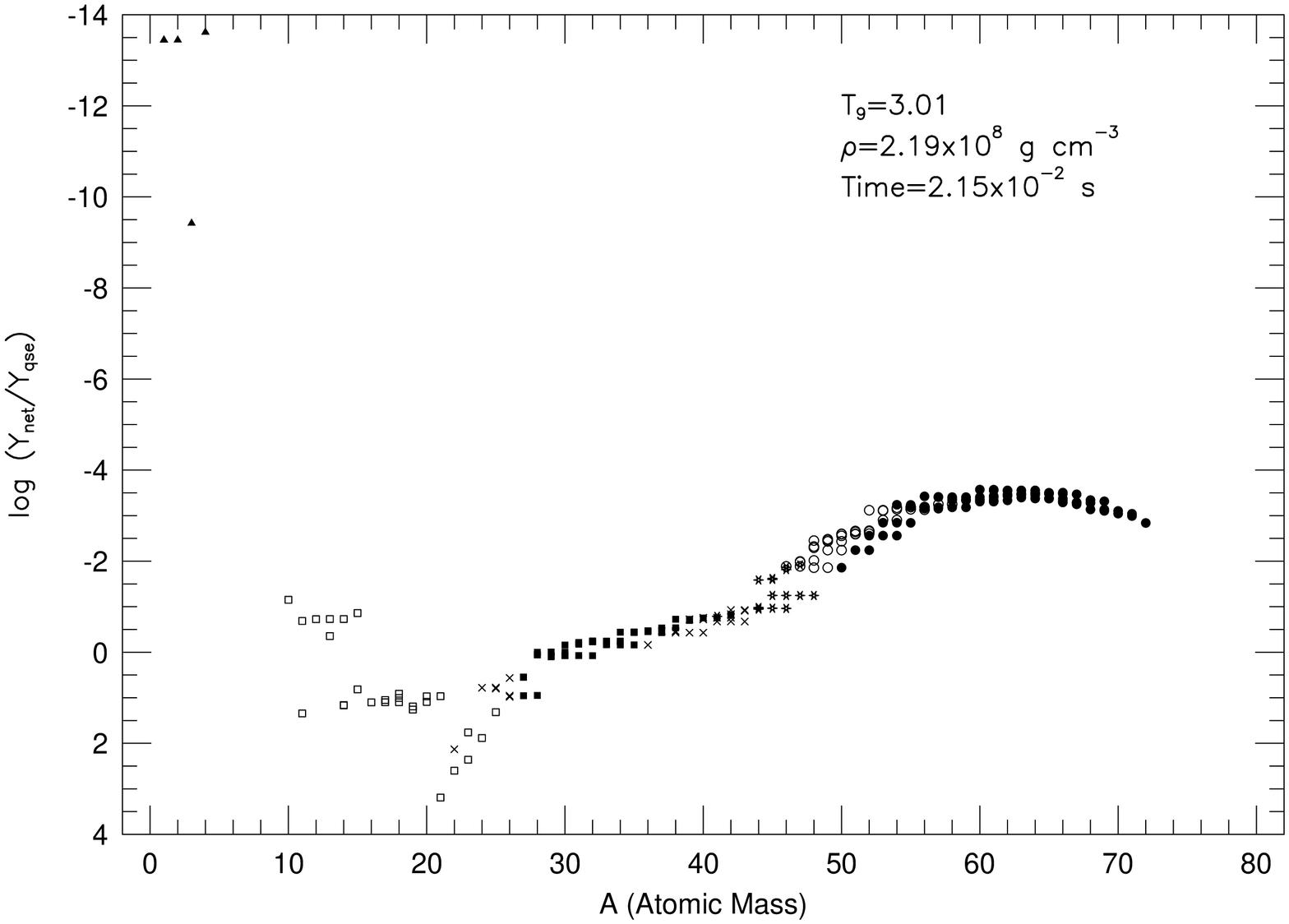}
    \caption{(d) Examination of the QSE group structure for $\tninei=5$, 
	$\rho_i= 10^9 \gcc$, and $Y_e=.48$ after elapsed time of $2.2 
	\times 10^{-2}$ seconds ($\tnine=3.0$, $\rho=2.2 \times 10^{8} \gcc$).  
	$Y_{net}$ are the network abundances, while $Y_{qse}$ are abundances 
	calculated assuming QSE from the network abundances of n, p, 
	\nuc{Si}{28}.}
    \label{fig:2d}
\end{figure}

Figure 2c follows the QSE group evolution further, showing the QSE group 
structure at an elapsed time of $1.5 \times 10^{-2}$ seconds, corresponding 
to $\tnine=3.5$ and $\rho= 3.5 \times 10^{8} \gcc$.  As the temperature 
continues to decline, the trend begun near $\tnine=4$ continues, with the 
light and iron peak QSE groups moving still further from quasi-equilibrium 
with the silicon group.  For $\tnine=3.5$, the spread from 
quasi-equilibrium among the group abundances is comparable to those for 
$\tnine=4.5$ and 4.  Columns 2 and 3 of Table~\ref{tab:ib} compare the 
network abundances with those predicted by membership in the appropriate 
QSE group.  The QSE abundances, $Y_{qse}$, in all three Tables in this 
article are calculated from the network abundances of free protons and 
neutrons and either \nuc{Si}{28} or \nuc{Ni}{56}.  Typically the 
discrepancy is a few \%, though somewhat larger at the edges of the groups.  
By $\tnine=3.25$, quasi-equilibrium is beginning to fail.  As a result the 
silicon and iron peak QSE groups are getting more ragged, with the 
abundances of the important nuclei within the silicon group having a 25\% 
spread from quasi-equilibrium and the spread within the iron peak group 
having grown to 10\% (ignoring nuclei with abundances smaller than 
$10^{-12}$).  As temperature and density continue to drop, more and more of 
the photodisintegration reactions which cause quasi-equilibrium freeze out, 
resulting in declining abundances of the free nucleons and \alp-particles.  
By the time $\tnine=3.0$ (elapsed time=$2.2 \times 10^{-2}$ seconds), the 
abundances of the light nuclei have dropped by more than an order of 
magnitude, severely skewing the QSE abundance pattern, which is shown in 
Figure 2d.  The silicon QSE group (which still comprises 10\% of the mass) 
displays a shear of almost an order of magnitude between \nuc{Si}{28} and 
neutron-rich Ar.  As comparison of the 4th and 5th columns of 
Table~\ref{tab:ib} reveals, important nuclei like \nuc{S}{34} and 
\nuc{Ar}{36} have abundances less than 50\% of that predicted by 
quasi-equilibrium with \nuc{Si}{28}.  Even some of the core nuclei of the 
iron peak, isotopes of Fe, Co, Ni, Cu and Zn, are displaced by as much as 
50\% from equilibrium with \nuc{Ni}{56}.  Most indicative of the breakdown 
of quasi-equilibrium, the abundance of \nuc{He}{4} is 30\% smaller than 
equilibrium with the free nucleons would require.  Globally, the breakdown 
in QSE also implies that the QSE abundances no longer satisfy the nucleon 
number conservation relation, $\sum A Y$.  While network convergence 
conditions require that $\sum AY_{net}$ change by less than one part in 
$10^{6}$ during a timestep, $\sum AY_{qse} \sim 1.3$ at this point.

\begin{table}[t]
    \centering
    \small 
    \caption{Selected abundances near freezeout for the \emph{incomplete 
    burning} example ($\tninei=5$, $\rho_i= 10^{9} \gcc$ and $Y_e=.48$)
    \label{tab:ib}}
    \begin{tabular}{c|ll|lll|l}
        \tableline
        Time (s)& \multicolumn{2} {|c|}{1.50\ttt{-2}} & \multicolumn{3} {c|} 
        {2.15\ttt{-2}} & \multicolumn{1}{c}{4.21\ttt{-2}} \\
        \tnine  & \multicolumn{2} {|c|}{3.51}  & \multicolumn{3} {c|}{3.01} & 
        \multicolumn{1}{c}{1.85} \\
        $\rho$ (\gcc) & \multicolumn{2} {|c|}{3.45\ttt{8}} & \multicolumn{3} 
        {c|}{2.18\ttt{8}} & \multicolumn{1}{c}{5.05\ttt{7}} \\
        \tableline \tableline
        Nucleus 
        &\multicolumn{1}{c}{$Y_{net}$}&\multicolumn{1}{c|}{$Y_{qse}$}&
        \multicolumn{1}{c}{$Y_{net}$}&\multicolumn{1}{c}{$Y_{qse}$}&
        \multicolumn{1}{c|}{$Y_{nse}$}&\multicolumn{1}{c}{$Y_{net}$} \\
        \tableline
        \nuc{n }{ } & 2.67\ttt{-13}& 2.67\ttt{-13}& 1.66\ttt{-15}& 1.66\ttt{-15}
        & 1.06\ttt{-15} & 8.68\ttt{-24} \\
        \nuc{p }{ } & 8.49\ttt{-9} & 8.49\ttt{-9} & 1.93\ttt{-10}& 1.93\ttt{-10}
        & 7.59\ttt{-10} & 4.76\ttt{-16} \\
        \nuc{He}{4} & 2.08\ttt{-8} & 1.90\ttt{-8} & 6.44\ttt{-10}& 9.56\ttt{-10}
        & 6.11\ttt{-9}  & 1.26\ttt{-14} \\
        \nuc{Si}{28}& 2.50\ttt{-3} & 2.50\ttt{-3} & 2.49\ttt{-3} & 2.49\ttt{-3} 
        & 4.08\ttt{-11} & 2.49\ttt{-3}  \\
        \nuc{Si}{30}& 3.69\ttt{-6} & 3.75\ttt{-6} & 2.12\ttt{-6} & 3.04\ttt{-6} 
        & 2.04\ttt{-14} & 1.72\ttt{-6}  \\
        \nuc{S}{32} & 8.03\ttt{-4} & 8.21\ttt{-4} & 8.24\ttt{-4} & 1.45\ttt{-3} 
        & 1.54\ttt{-10} & 8.27\ttt{-4}  \\
        \nuc{S}{34} & 2.87\ttt{-5} & 3.10\ttt{-5} & 2.76\ttt{-5} & 7.58\ttt{-5} 
        & 3.30\ttt{-12} & 2.66\ttt{-5}  \\
        \nuc{Ar}{36}& 1.00\ttt{-4} & 1.07\ttt{-4} & 9.67\ttt{-5} & 2.83\ttt{-4} 
        & 1.95\ttt{-10} & 9.74\ttt{-5}  \\
        \nuc{Ar}{38}& 2.22\ttt{-5} & 2.59\ttt{-5} & 2.43\ttt{-5} & 1.29\ttt{-4} 
        & 3.66\ttt{-11} & 2.42\ttt{-5}  \\
        \nuc{Ca}{40}& 4.83\ttt{-5} & 5.62\ttt{-5} & 4.88\ttt{-5} & 2.79\ttt{-4} 
        & 1.25\ttt{-9}  & 4.92\ttt{-5}  \\
        \nuc{Ca}{42}& 8.22\ttt{-7} & 1.06\ttt{-6} & 7.51\ttt{-7} & 6.35\ttt{-6} 
        & 1.17\ttt{-11} & 7.50\ttt{-7}  \\
        \nuc{Ti}{46}& 4.56\ttt{-6} & 3.22\ttt{-6} & 4.12\ttt{-6} & 2.24\ttt{-7} 
        & 3.83\ttt{-9}  & 4.13\ttt{-6}  \\
        \nuc{Ti}{48}& 5.55\ttt{-7} & 4.86\ttt{-7} & 3.05\ttt{-7} & 6.09\ttt{-8} 
        & 4.27\ttt{-10} & 2.78\ttt{-7}  \\
        \nuc{Fe}{54}& 1.09\ttt{-2} & 1.09\ttt{-2} & 1.13\ttt{-2} & 1.38\ttt{-2} 
        & 9.59\ttt{-3}  & 1.13\ttt{-2}  \\
        \nuc{Fe}{56}& 2.12\ttt{-3} & 2.13\ttt{-3} & 2.35\ttt{-3} & 4.39\ttt{-3} 
        & 1.26\ttt{-3}  & 2.37\ttt{-3}  \\
        \nuc{Fe}{58}& 3.23\ttt{-8} & 3.21\ttt{-8} & 1.61\ttt{-8} & 2.47\ttt{-8} 
        & 2.90\ttt{-9}  & 1.50\ttt{-8}  \\
        \nuc{Ni}{56}& 7.02\ttt{-6} & 7.02\ttt{-6} & 2.33\ttt{-6} & 2.33\ttt{-6} 
        & 2.53\ttt{-5}  & 2.06\ttt{-6}  \\
        \nuc{Ni}{58}& 7.67\ttt{-4} & 7.65\ttt{-4} & 7.12\ttt{-4} & 1.28\ttt{-3} 
        & 5.69\ttt{-3}  & 7.14\ttt{-4}  \\
        \nuc{Ni}{60}& 8.67\ttt{-5} & 8.45\ttt{-5} & 8.74\ttt{-5} & 2.29\ttt{-4} 
        & 4.18\ttt{-4}  & 8.75\ttt{-5}  \\
        \nuc{Zn}{60}& 3.10\ttt{-12}& 3.09\ttt{-12}& 9.40\ttt{-14}& 1.68\ttt{-13}
        & 1.16\ttt{-11} & 2.48\ttt{-15} \\
        \nuc{Zn}{62}& 3.09\ttt{-9} & 3.00\ttt{-9} & 4.66\ttt{-10}& 1.18\ttt{-9} 
        & 3.34\ttt{-8}  & 2.49\ttt{-10} \\
        \nuc{Zn}{64}& 2.32\ttt{-9} & 2.18\ttt{-9} & 8.78\ttt{-10}& 1.97\ttt{-9} 
        & 2.30\ttt{-8}  & 8.27\ttt{-10} \\
        \tableline
    \end{tabular}
%    \vspace{2.5in}
\end{table}

Further decline in temperature and density results in further reductions in 
the light nuclear abundances.  As comparison of columns 4 and 7 of 
Table~\ref{tab:ib} reveals, between $2.2 \times 10^{-2}$ and $4.2 \times 
10^{-2}$ seconds, corresponding to $\tnine=3.0$ and $\tnine=1.85$, the 
abundances of the free protons and \alp-particles drop by roughly five more 
orders of magnitude, with the free neutron abundance dropping by more than 
8 more order of magnitude.  While these continued captures remove any hint 
of quasi-equilibrium with the current free nucleon abundances, further 
examination of columns 4 and 7 of Table~\ref{tab:ib} reveals that the most 
important products (those with mass fractions greater than .1\%) have 
undergone only small changes in abundance.  Even though quasi-equilibrium 
has become suspect by $\tnine=3.0$, abundances calculated assuming QSE
(using the network values of the abundances of \nuc{Si}{28}, \nuc{Ni}{56},
free protons and neutrons), at 
this temperature (column 5 of Table~\ref{tab:ib}) are a better estimate of 
the final abundances than those predicted by the freezeout of NSE at 
$\tnine =3.0$ (shown in column 6 of Table~\ref{tab:ib}).  While this is not 
surprising for a case which fails to exhaust silicon, it is still worth 
noting that the relative abundances among the dominant iron peak nuclei are 
also better predicted by QSE than NSE. As we discussed in HT96, QSE and the 
greater binding energy of the neutron rich members of the iron peak group 
result in preferential storage of neutrons in the iron peak group.  In this 
case, even though the iron peak group contains 90\% of the mass fraction, 
comparison of the network abundances at \tnine=3 (column 5) to the 
corresponding NSE abundances (column 7) reveals significant enhancement of 
the neutron-rich iron peak nuclei.  Such enhancements are better estimated 
by QSE than NSE. For example, the ratio of the QSE abundances of 
\nuc{Fe}{56} to \nuc{Fe}{54} frozen at \tnine=3.5 is .20.  This is much 
closer to the final network ratio $Y(\nuc{Fe}{56})/Y(\nuc{Fe}{54})=.21$ 
than that predicted by NSE frozen around \tnine=3 (.14).

If our hope to in the future use QSE to reduce the computational cost of 
silicon burning nucleosynthesis is to be realized, we must understand where 
we can safely employ QSE. Clearly, the sharp drop in the free nucleon 
abundances, and the attendant departure from QSE, seen in this case below 
\tnine=3.0, can not be followed by a method employing QSE. However, since 
abundances other than the free nucleons and \alp-particles are little 
affected, we can provide reasonable estimates for the final abundances by 
freezing the QSE abundances prior to the actual freezeout of 
photodisintegrations.  For example, comparison of columns 3 and 7 of 
Table~\ref{tab:ib}, reveals that the QSE abundance, frozen at $\tnine=3.5$, 
provides good $(\pm10\%)$ estimates of the final nuclear abundances for all 
of the abundances larger than $10^{-5}$.  Thus we conclude that 
quasi-equilibrium and, by extension, future methods which employ 
quasi-equilibrium, can provide good abundance estimates for cases of 
silicon burning which fail to reach silicon exhaustion and that these 
estimates are significantly better than those based on NSE.

\section{Normal Freezeout}

The next test of the applicability of quasi-equilibrium is that of complete 
silicon burning and the {\it normal freezeout}, where previous authors have 
indicated that the nuclei within the mass zone reach NSE and then remain in 
NSE as the mass zone cools.  To this end, we present an example with 
$\tninei =6$, $\rho=10^9 \gcc$, and $Y_e=.48$.  Early in the calculation, 
after an elapsed time of $2.0 \times 10^{-8}$ seconds, the QSE group 
structure shows clear evidence of three distinct groups, with the 
abundances of the iron peak and light QSE groups being one and ten orders 
of magnitude smaller, respectively, than equilibrium with the silicon QSE 
group would require.  As in the previous example, these groups initially 
converge toward mutual equilibrium.  By the time $5.1 \times 10^{-7}$ 
seconds have passed, with 75\% of the mass still concentrated within the 
original silicon group, the iron peak and silicon groups reach mutual 
equilibrium.

\begin{figure}[t]
    \centering
    \includegraphics[angle=90,width=\textwidth]{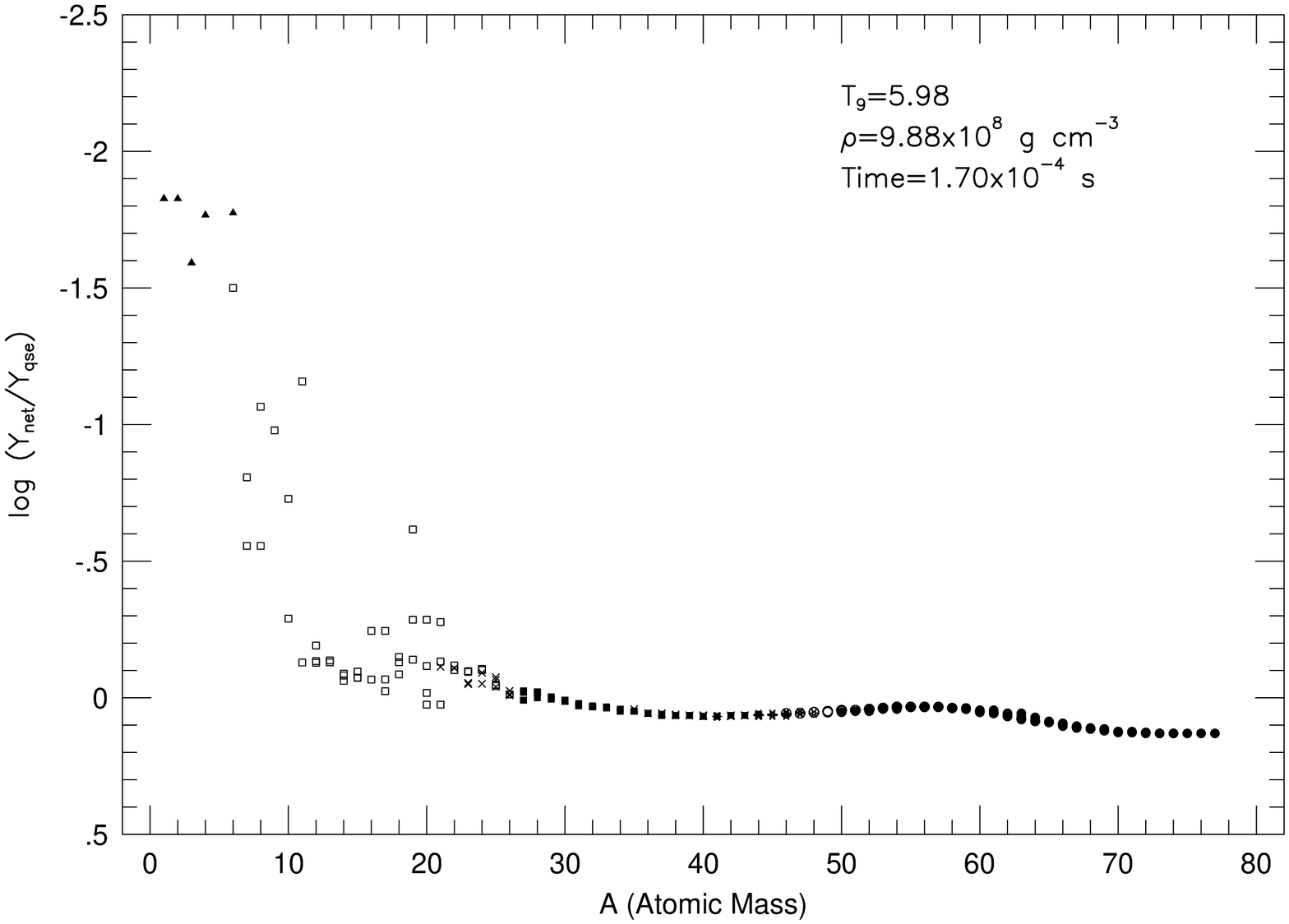}
    \caption{(a) Examination of the QSE group structure for $\tninei=6$, 
	$\rho_i=10^9 \gcc$, and $Y_e=.48$, after elapsed times of $1.70 
	\times 10^{-4}$ seconds ($\tnine=5.98$ and $\rho=9.88 \times 10^{8} 
	\gcc$).  $Y_{net}$ are the network abundances, while $Y_{qse}$ are 
	abundances calculated assuming QSE from the network abundances of n, p, 
	\nuc{Si}{28}.}
    \label{fig:3a}
    \addtocounter{figure}{-1}
\end{figure}

\begin{figure}[ht]
    \centering
    \includegraphics[angle=90,width=\textwidth]{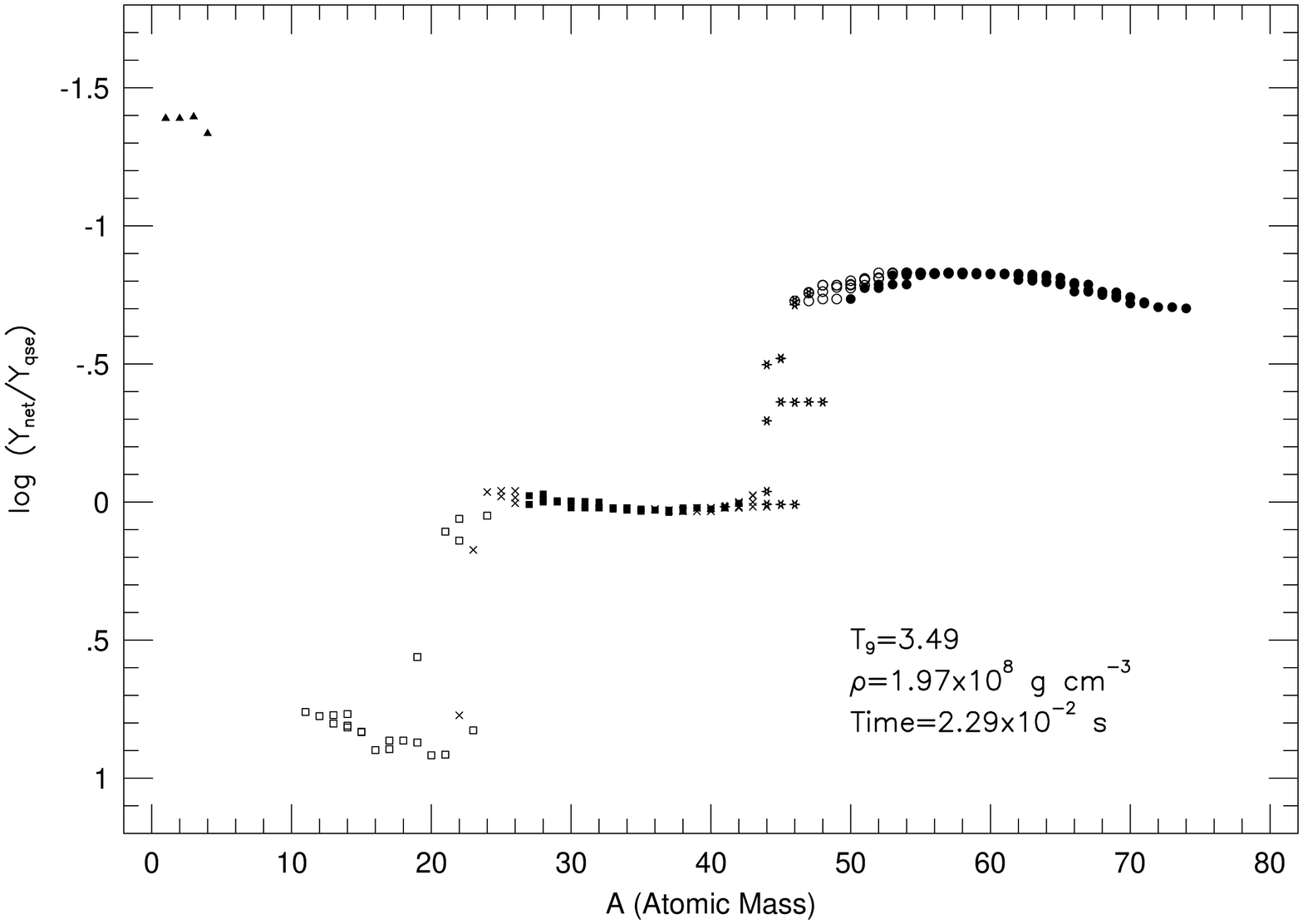}
    \caption{(b) Examination of the QSE group structure for $\tninei=6$,
	$\rho_i= 10^9 \gcc$, and $Y_e=.48$, after elapsed time of $2.29 \times 
        10^{-2}$ seconds ($\tnine=3.49$ and $\rho=1.97 \times 10^{8}\gcc$).  
        $Y_{net}$ are the network abundances, while $Y_{qse}$ are abundances 
        calculated assuming QSE from the network abundances of n, p, 
	\nuc{Si}{28}.}
    \label{fig:3b}
\end{figure}

After an elapsed time of $1.7 \times 10^{-4}$ seconds, the temperature has 
dropped to $\tnine=5.98$, with a density of $9.9 \times 10^{8} \gcc$.  As 
Figure 3a demonstrates, the convergence of the QSE groups has continued, 
with the abundances of the light group now greater than 1\% of their 
silicon quasi-equilibrium abundance.  The combined silicon and iron peak 
QSE group now stretches as far as A=10, with the abundances of \nuc{C}{12}, 
\nuc{O}{16}, and \nuc{Ne}{20} all being within 25\% of quasi-equilibrium 
with \nuc{Si}{28}.  The deviations near A=19 are due to an unbalanced 
reaction in the reaction rate library.  While this error effects 
\nuc{F}{19} and, to a lesser extent, several of its neighbors with trace 
abundances, it is unimportant to the dominat flows which link the light QSE 
group to the silicon QSE group.  Though the members of this large QSE group 
dominate the mass fraction, the absence of the free nucleons from the 
unified QSE group indicates that NSE is not yet fully established.  Figure 
4a compares the network abundances at this point in the calculation to the 
abundances that NSE predicts for these conditions.  For Figures 4 and 6, 
the filled shapes represent network abundances larger than $10^{-6}$, and 
the hollow shapes denote abundances less than $10^{-12}$.  The sided-ness 
of the shapes still identifies their QSE group membership as in Figures 2, 
3 \& 5.  By this time ($1.7 \ttt{-4}$ s), the dominant members of the iron 
peak (which now contain almost 90\% of the mass) and the light nuclei 
(which represent roughly 0.3\% of the mass) are very close to NSE. However 
the nuclei from carbon through silicon, though in quasi-equilibrium, lag 
behind with abundances as much as an order of magnitude larger than NSE 
would predict.

\begin{figure}[!t]
    \centering
    \includegraphics[angle=90,width=\textwidth,clip]{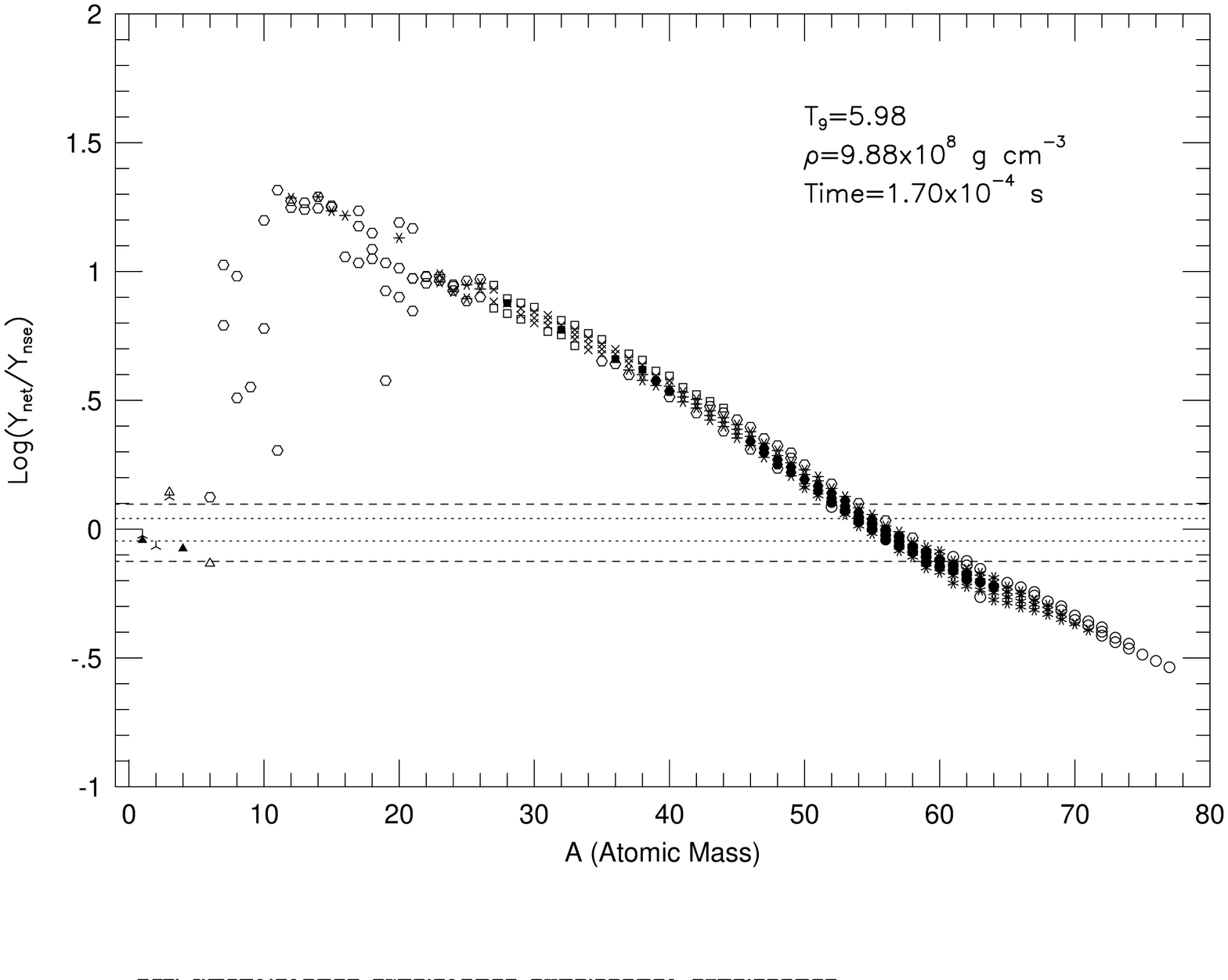}
    \caption{(a) Comparison of the network abundances to the NSE abundances 
	for a case with $\tninei=6$, $\rho_i=10^9 \gcc$, and $Y_e=.48$, 
	after elapsed times of $1.7 \times 10^{-4}$ seconds ($\tnine=5.98$ 
	and $\rho=9.9 \times 10^{8} \gcc$).  The filled shapes represent 
	network abundances larger than $10^{-6}$, and the hollow shapes 
	denote abundances less than $10^{-12}$.}
    \label{fig:4a}
    \addtocounter{figure}{-1}
\end{figure}

\begin{figure}[!t]
    \centering
    \includegraphics[angle=90,width=\textwidth,clip]{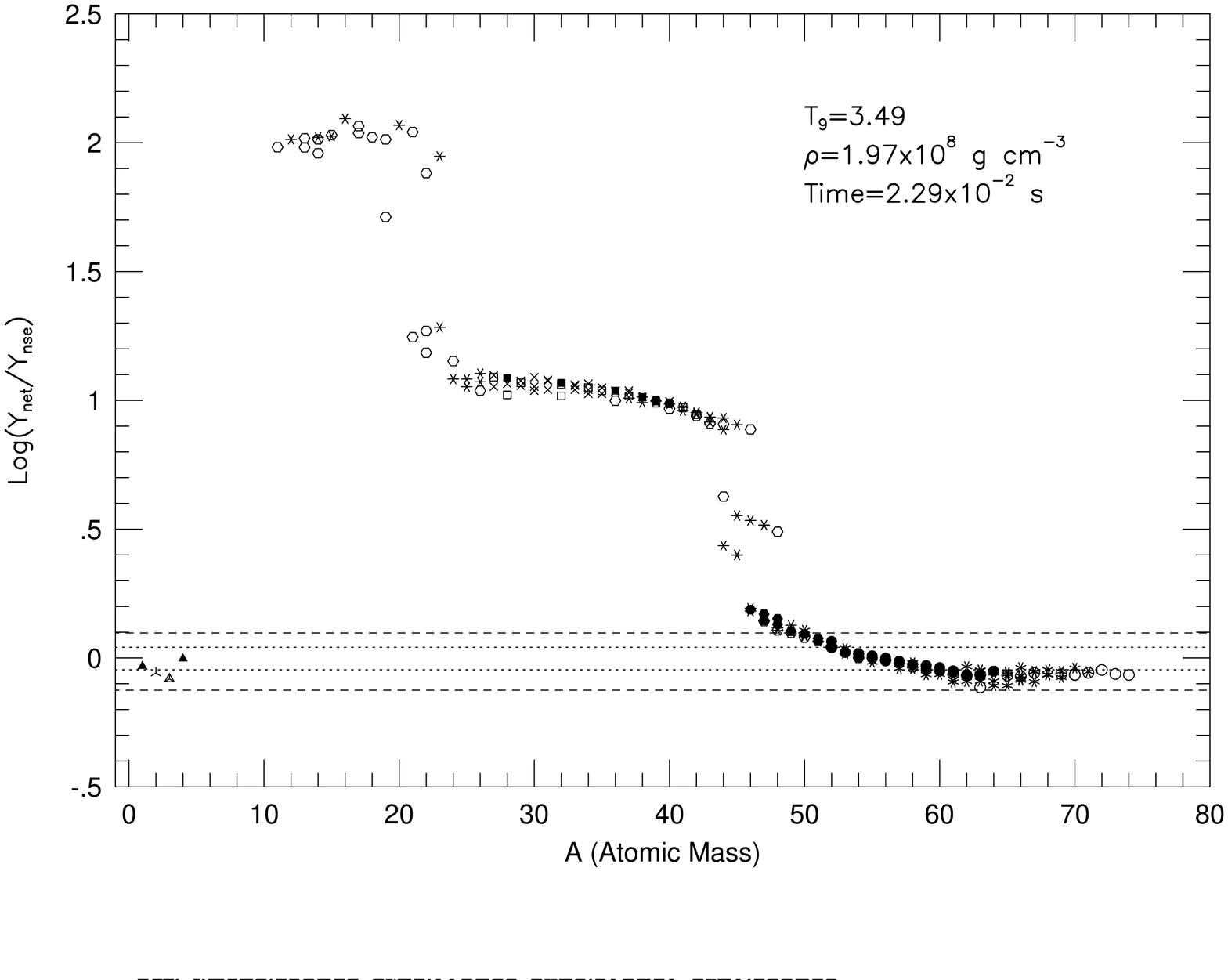}
    \caption{(b) Comparison of the network abundances to the NSE abundances 
	for a case with $\tninei=6$, $\rho_i=10^9 \gcc$, and $Y_e=.48$, 
	after elapsed time of $2.3 \times 10^{-2}$ seconds ($\tnine=3.5$ 
	and $\rho=1.97 \times 10^{8} \gcc$.  The filled shapes represent 
	network abundances larger than $10^{-6}$, and the hollow shapes 
	denote abundances less than $10^{-12}$.}
    \label{fig:4b}
\end{figure}

As the calculation continues, these laggards converge further toward the
current NSE abundance pattern.  However, significant discrepancies remain 
between the network abundances and those predicted by NSE for the 
intermediate mass nuclei.  For example, by an elapsed time of $1.5 \times 
10^{-3}$ seconds ($\tnine=5.8$ and $\rho= 9.0 \times 10^{8} \gcc$), 
\nuc{Si}{28} is nearly 3 times more abundant than it would be if NSE were 
completely established.  Quasi-equilibrium provides a better estimate for 
many nuclei, with the single QSE group stretching from carbon to germanium 
displaying deviations from quasi-equilibrium of less than 25\%.  As our 
hypothetical hydrodynamic mass zone continues to expand and cool, the 
abundance distribution never catches up with NSE. By an elapsed time of 
$1.7 \times 10^{-2}$ seconds, with \tnine\ having dropped to 4.0 and $\rho$ 
to $3.0 \times 10^{8} \gcc$, the network abundances for the intermediate 
mass elements are not significantly closer to those calculated from NSE, 
with \nuc{Si}{28} remaining 2.5 times more abundant than predicted by NSE. 
The single large QSE group has also begun to fragment.  By the time the 
temperature has dropped to $\tnine=3.5$ (elapsed time $= 2.29 \times 
10^{-2}$ seconds), there are 4 distinct QSE groups.  As Figure 3b shows, 
the silicon and iron groups are again distinct, separated by a fringe of 
nuclei; \nuc{K}{45}, \nuc{Ca}{44}, \nuc{Ca}{45}, \nuc{Sc}{45}, 
\nuc{Ti}{45}, \nuc{V}{46}, \nuc{Cr}{47}, and \nuc{Mn}{48}.  These nuclei 
are familiar from the analysis presented in HT96.  As the QSE groups 
separate, the fracture occurs where they were joined.  Comparison of 
columns 2 and 3 of Table \ref{tab:nf} reveals that within these groups, the 
abundances predicted by QSE differ by only a few \% from the network 
abundances.  In addition to these groups and the light group, the nuclei in 
the C-O-Ne region are within 25\% of mutual equilibrium.  These nuclei are 
almost an order of magnitude overabundant for quasi-equilibrium with the 
silicon group.  Comparison between the network abundances and those 
predicted by NSE, shown in Figure 4b, reveals larger discrepancies, with 
the C-O-Ne nuclei roughly 2 orders of magnitude more abundant than NSE 
would predict and the silicon group an order of magnitude overabundant.  
The abundances within the light group and the iron peak group, which, now 
contains more than 99\% of the mass, are quite close to their NSE values.

In this case, like that discussed in the previous section, in the vicinity 
of \tnine = 3.0, quasi-equilibrium breaks down.  Though the discrepancies 
between the network and QSE abundances (which are based on the network 
abundances of n, p, \nuc{Si}{28} and \nuc{Ni}{56}) at \tnine=3.5 (columns 
2 and 3 of 
Table \ref{tab:nf}) are small, examination of columns 4 and 5 of Table 
\ref{tab:nf} reveals much larger differences by the time \tnine=3.0.  Among 
the nuclei which dominate the mass fraction, slightly neutron rich members 
of the iron group, predictions based on QSE err by as much as 50\%.  It is 
the decline in the abundances of the light nuclei, which drop by more than 
an order of of magnitude between \tnine=3.5 and \tnine=3.0, which cause 
this discrepancy.  Comparison of the abundances of the dominant members of 
the iron peak group with abundance predictions based on NSE (column 6 of 
Table \ref{tab:nf}), confirms that NSE does provide a good estimate of the 
iron group (and light group) abundances.  However the differences between 
the network and NSE for intermediate nuclei persist.  For example, 
\nuc{Ca}{40}, the most abundant of the intermediate mass elements with a 
mass fraction of $9\ttt{-6}$, is 160 times more abundant than NSE would 
predict at \tnine=3.0.  As Column 7 of Table \ref{tab:nf} reveals, these 
over abundances of the intermediate mass elements persist once the final 
freezeout occurs.  This is consistent with the results of \cite{MeKC96}, 
who found that, for significantly more neutronized material, such high 
density (low entropy) cases can result in final abundances of \nuc{Ca}{48} 
as large as .01, millions of times what NSE, frozen at \tnine=3.5, would 
predict.  Comparison of columns 7 and 4 also reveals that in this case, 
like the previous, the abundances of the light group nuclei are the only 
ones which change markedly between \tnine= 3.0 and \tnine=1.8.

\begin{table}[t]
    \centering
    \small
    \caption{Selected abundances near freezeout for the \emph{normal freezeout} 
    example ($\tninei=6$, $\rho_i= 10^{9} \gcc$ and $Y_e=.48$)
    \label{tab:nf}}
    \begin{tabular}{c|ll|lll|l}
        \tableline
        Time (s)& \multicolumn{2} {|c|}{2.29\ttt{-2}} & \multicolumn{3} {c|} 
        {2.93\ttt{-2}} & \multicolumn{1}{c}{5.10\ttt{-2}} \\
        \tnine  & \multicolumn{2} {|c|}{3.49}  & \multicolumn{3} {c|}{3.00} & 
        \multicolumn{1}{c}{1.80} \\
        $\rho$ (\gcc)& \multicolumn{2} {|c|}{1.97\ttt{8}} & \multicolumn{3} 
        {c|}{1.25\ttt{8}} & \multicolumn{1}{c}{2.68\ttt{7}} \\
        \tableline \tableline
        Nucleus &\multicolumn{1}{c}{$Y_{net}$}&\multicolumn{1}{c|}{$Y_{qse}$}&
        \multicolumn{1}{c}{$Y_{net}$}&\multicolumn{1}{c}{$Y_{qse}$}&
        \multicolumn{1}{c|}{$Y_{nse}$}&\multicolumn{1}{c}{$Y_{net}$} \\
        \tableline
        \nuc{n }{ } & 3.26\ttt{-13}& 3.26\ttt{-13}& 1.96\ttt{-15}& 1.96\ttt{-15}
        & 1.68\ttt{-15}& 3.74\ttt{-24} \\
        \nuc{p }{ } & 5.01\ttt{-8} & 5.01\ttt{-8} & 1.35\ttt{-9} & 1.35\ttt{-9} 
        & 1.30\ttt{-9} & 2.65\ttt{-15} \\
        \nuc{He}{4} & 2.93\ttt{-7} & 2.58\ttt{-7} & 1.34\ttt{-8} & 1.57\ttt{-8} 
        & 1.07\ttt{-8} & 3.10\ttt{-10}\\
        \nuc{Si}{28}& 2.42\ttt{-8} & 2.42\ttt{-8} & 1.82\ttt{-8} & 1.82\ttt{-8} 
        & 6.35\ttt{-11}& 1.76\ttt{-8} \\
        \nuc{Si}{30}& 2.34\ttt{-11}& 2.23\ttt{-11}& 9.70\ttt{-12}& 1.20\ttt{-11}
        & 3.10\ttt{-14}& 6.47\ttt{-12}\\
        \nuc{S}{32} & 6.03\ttt{-8} & 5.76\ttt{-8} & 5.55\ttt{-8} & 9.15\ttt{-8} 
        & 2.18\ttt{-10}& 5.55\ttt{-8} \\
        \nuc{S}{34} & 1.45\ttt{-9} & 1.35\ttt{-9} & 1.14\ttt{-9} & 2.61\ttt{-9} 
        & 4.59\ttt{-12}& 9.64\ttt{-10}\\
        \nuc{Ar}{36}& 5.71\ttt{-8} & 5.34\ttt{-8} & 5.58\ttt{-8} & 1.50\ttt{-7} 
        & 2.43\ttt{-10}& 5.64\ttt{-8} \\
        \nuc{Ar}{38}& 8.43\ttt{-9} & 8.04\ttt{-9} & 7.52\ttt{-9} & 3.77\ttt{-8} 
        & 4.52\ttt{-11}& 6.67\ttt{-9} \\
        \nuc{Ca}{40}& 2.04\ttt{-7} & 1.95\ttt{-7} & 2.18\ttt{-7} & 1.22\ttt{-6} 
        & 1.36\ttt{-9} & 2.20\ttt{-7} \\
        \nuc{Ca}{42}& 2.29\ttt{-9} & 2.29\ttt{-9} & 2.12\ttt{-9} & 1.52\ttt{-8} 
        & 1.25\ttt{-11}& 2.13\ttt{-9} \\
        \nuc{Ti}{46}& 5.08\ttt{-8} & 4.05\ttt{-8} & 3.59\ttt{-8} & 1.95\ttt{-9} 
        & 3.49\ttt{-9} & 3.62\ttt{-8} \\
        \nuc{Ti}{48}& 4.21\ttt{-9} & 3.82\ttt{-9} & 1.33\ttt{-9} & 2.92\ttt{-10}
        & 3.85\ttt{-10}& 9.04\ttt{-10}\\
        \nuc{Fe}{54}& 9.53\ttt{-3} & 9.58\ttt{-3} & 9.88\ttt{-3} & 1.17\ttt{-2} 
        & 9.73\ttt{-3} & 9.93\ttt{-3} \\
        \nuc{Fe}{56}& 1.15\ttt{-3} & 1.16\ttt{-3} & 1.22\ttt{-3} & 2.04\ttt{-3} 
        & 1.26\ttt{-3} & 1.22\ttt{-3} \\
        \nuc{Fe}{58}& 1.05\ttt{-8} & 1.06\ttt{-8} & 3.84\ttt{-9} & 6.17\ttt{-9} 
        & 2.80\ttt{-9} & 2.99\ttt{-9} \\
        \nuc{Ni}{56}& 7.29\ttt{-5} & 7.29\ttt{-5} & 3.19\ttt{-5} & 3.19\ttt{-5} 
        & 2.46\ttt{-5} & 2.90\ttt{-5} \\
        \nuc{Ni}{58}& 5.04\ttt{-3} & 5.07\ttt{-3} & 5.30\ttt{-3} & 9.77\ttt{-3} 
        & 5.56\ttt{-3} & 5.34\ttt{-3} \\
        \nuc{Ni}{60}& 3.49\ttt{-4} & 3.49\ttt{-4} & 3.57\ttt{-4} & 9.61\ttt{-4} 
        & 4.04\ttt{-4} & 3.58\ttt{-4} \\
        \nuc{Zn}{60}& 2.10\ttt{-10}& 2.11\ttt{-10}& 9.96\ttt{-12}& 1.82\ttt{-11}
        & 9.59\ttt{-12}& 4.48\ttt{-13}\\
        \nuc{Zn}{62}& 1.33\ttt{-7} & 1.32\ttt{-7} & 2.84\ttt{-8} & 7.16\ttt{-8} 
        & 2.78\ttt{-8} & 1.72\ttt{-8} \\
        \nuc{Zn}{64}& 6.47\ttt{-8} & 6.02\ttt{-8} & 3.01\ttt{-8} & 6.63\ttt{-8} 
        & 1.90\ttt{-8} & 2.89\ttt{-8} \\
        \tableline
    \end{tabular}
\end{table}

As in the previous example, comparison of the network abundances at 
freezeout with those corresponding to \tnine=3.5 reveals relatively little 
change.  As a result, abundances predicted assuming QSE but frozen around 
\tnine=3.5 
provide a good estimate for the final abundances of all nuclei save the 
lightest.  As comparison of columns 3 and 7 of Table~\ref{tab:nf} reveals, 
for the most abundant species (those with mass fractions greater than .002) 
and many other species, this estimate is within 5\% of the final abundance.  
Among the iron peak nuclei which dominate the mass fraction, these 
abundance predictions are comparable to those based on NSE while among the 
intermediate mass elements they are better by more than an order of 
magnitude.  As we discussed earlier, once the principal nucleus of a QSE 
group reaches equilibrium with the free nucleons, the equations of 
quasi-equilibrium are identical to those of NSE. Thus it is not surprising 
that quasi-equilibrium can provide estimates of abundances that agree well 
with NSE. However, with this example we have demonstrated that 
quasi-equilibrium can provide a better estimate of the abundances of 
individual species than NSE when the network is chasing the receding target 
of complete equilibrium.  Thus analysis of this example and others which 
result in normal freezeout indicate the quasi-equilibrium does provide an 
excellent estimation of the abundances of many species, when silicon is 
exhausted at high density or low entropy.  With the small changes in the 
dominant abundances seen following freezeout of the QSE, methods which 
employ QSE to reduce the size of the nuclear network should prove reliable 
for calculation of normal freezeout.

\section{\alp-rich Freezeout}

The final test is then to examine the case of silicon exhaustion at low 
density, which previous authors have dubbed \emph{\alp-rich freezeout}.  We 
consider an example of our model for explosive silicon burning, with 
$\tninei=6$, $\rho_i=10^{7} \gcc$, and $Y_e=.498$.  While at early times 
the silicon and iron peak groups are separate, by an elapsed time of $2.9 
\times 10^{-7}$ seconds, the abundances of these groups are with 15\% of 
mutual equilibrium.  While significant nucleosynthesis has occurred to this 
point, silicon is far from exhaustion, with a silicon group mass fraction 
of 65\% compared to 11\% of the mass being found among the iron peak 
nuclei.  As Figure 5a shows, the light QSE group, which at this point 
contains more than 13\% of the nuclear mass fraction, is more than 5 orders 
of magnitude underabundant for equilibrium with this combined QSE group.  
By the time $1.5 \times 10^{-4}$ seconds have elapsed, the light group, 
which now represents 36\% of the mass, has converged considerably toward 
equilibrium with this large QSE group, with the network abundance of the 
light nuclei almost 30\% of that required for quasi-equilibrium with 
\nuc{Si}{28}.  The combined silicon and iron peak group is extremely well 
equilibrated at this time, with only a 4\% spread separating germanium from 
quasi-equilibrium with silicon.  Comparison with abundances predicted by 
NSE, shown in Fig.  6a, reveals that these nuclei are also reasonably close 
to NSE, with the network abundance of \nuc{Si}{28} being twice its NSE 
value.  The nuclei of the light and Si-Fe QSE groups do show a relatively 
large spread in abundance when compared to NSE. This spread and the tilt 
downward with increasing atomic mass shown in this figure are ascribable to 
a slight underabundance of free protons and \alp-particles.

\begin{figure}[t]
    \centering
    \includegraphics[angle=90,width=\textwidth]{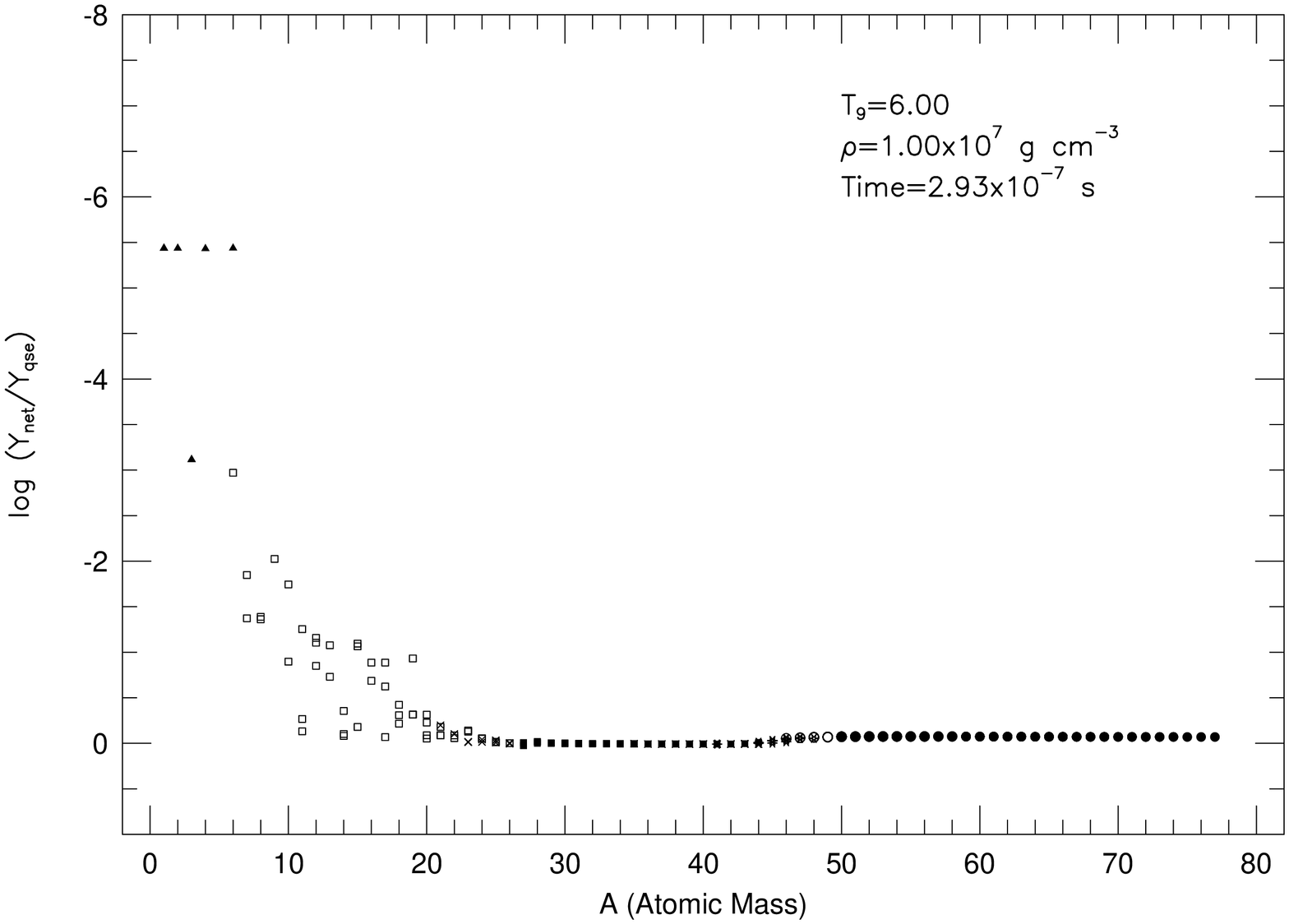}
    \caption{(a) Examination of the QSE group structure for $\tninei=6$, 
       $\rho_i= 10^7 \gcc$, and $Y_e=.498$, after elapsed time of $2.9 \times 
        10^{-7}$ seconds.  $Y_{net}$ are the network abundances, while 
	$Y_{qse}$ are abundances calculated assuming QSE from the network 
	abundances of n, p, \nuc{Si}{28}.}
    \label{fig:5a}
    \addtocounter{figure}{-1}
\end{figure}

\begin{figure}[ht]
    \centering
    \includegraphics[angle=90,width=\textwidth]{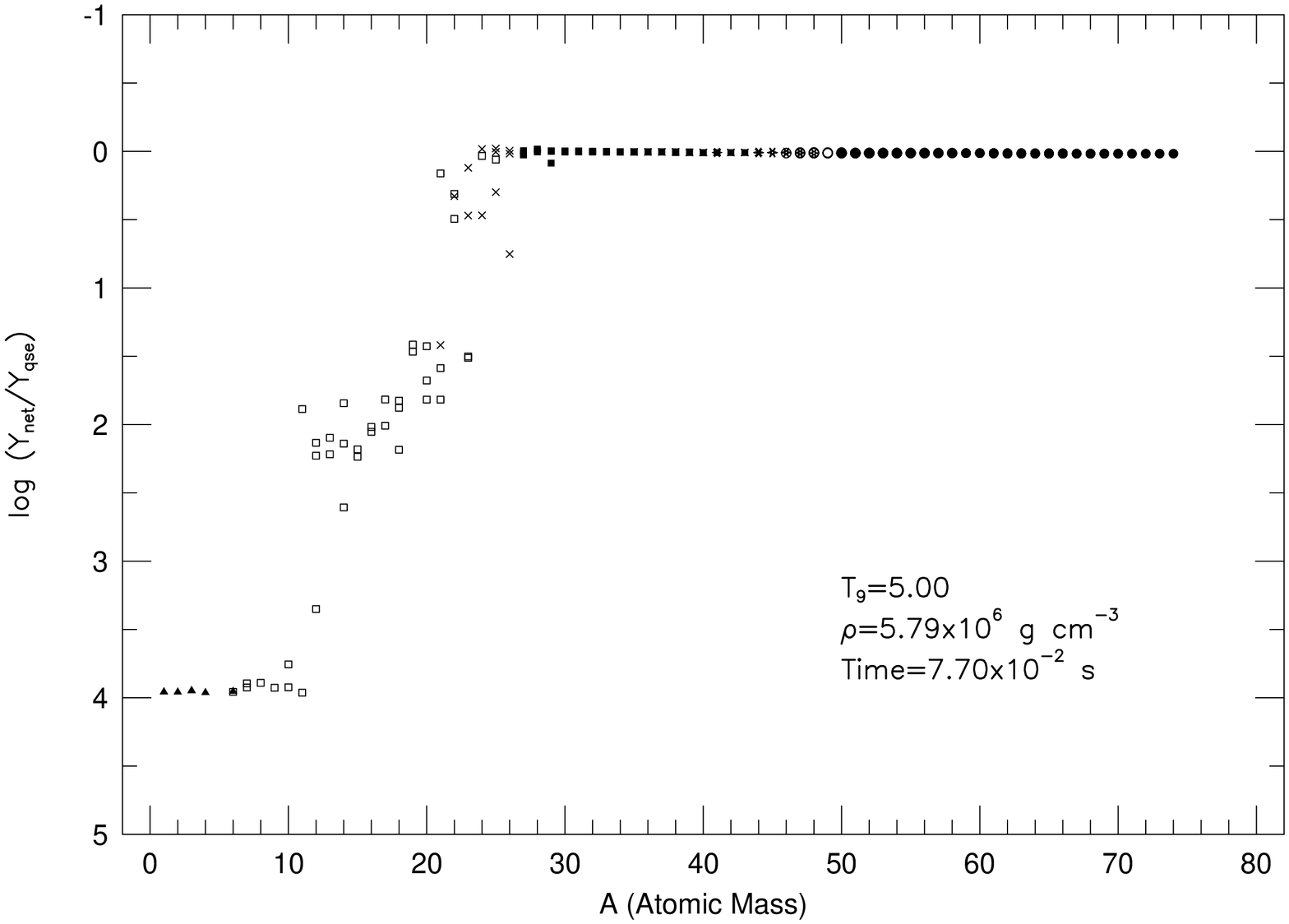}
    \caption{ (b) Examination of the QSE group structure for $\tninei=6$, 
         $\rho_i=10^7 \gcc$, and $Y_e=.498$, after elapsed time of $7.70 \times 
        10^{-2}$ seconds ($\tnine=5.0$ and $\rho=5.8 \times 10^{6} \gcc$).  
        $Y_{net}$ are the network abundances, while $Y_{qse}$ are abundances 
        calculated assuming QSE from the network abundances of n, p, 
	\nuc{Si}{28}.}
    \label{fig:5b}
    \addtocounter{figure}{-1}
\end{figure}

\begin{figure}[ht]
    \centering
    \includegraphics[angle=90,width=\textwidth]{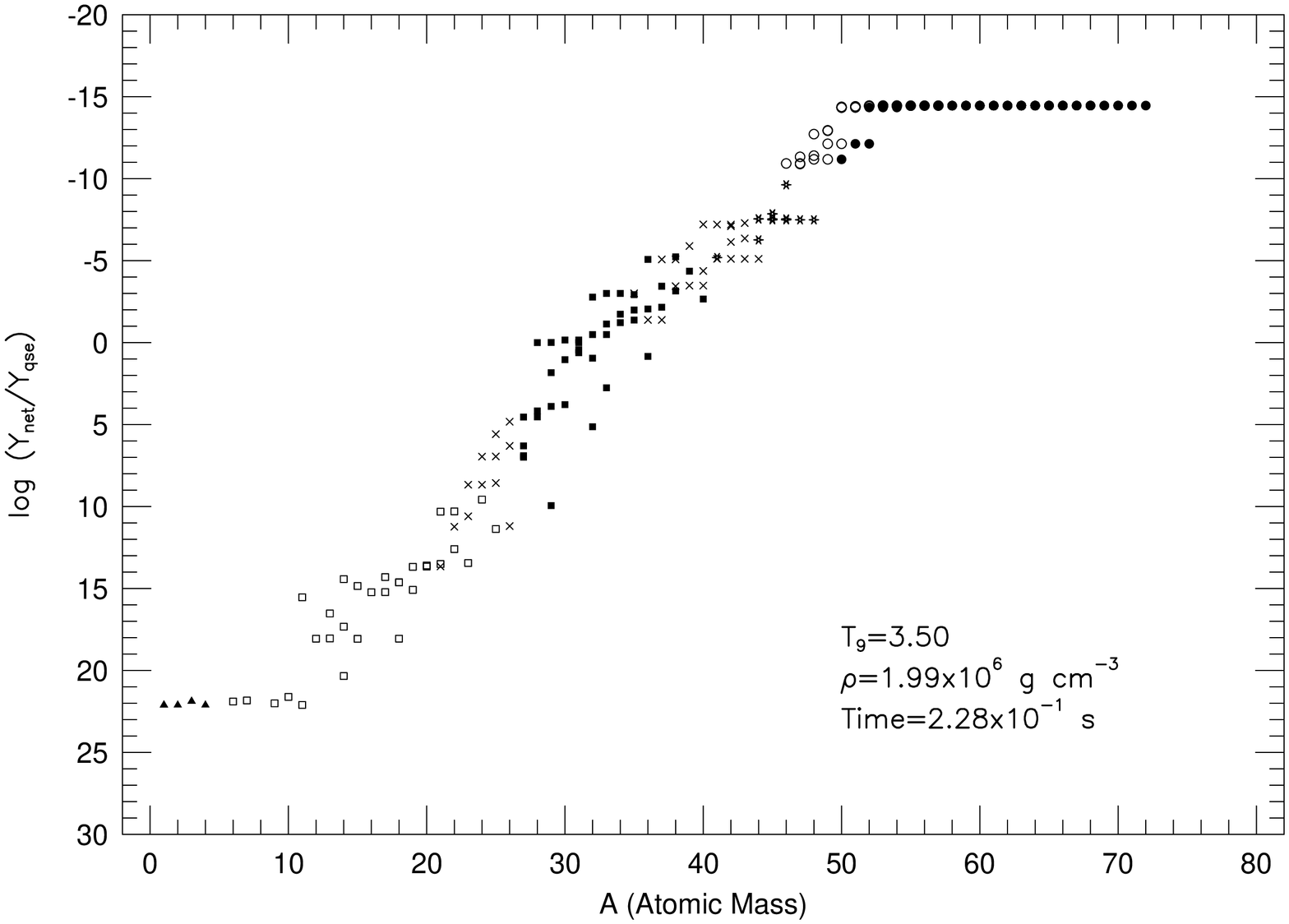}
    \caption{(c) Examination of the QSE group structure for $\tninei=6$, 
        $\rho_i=10^7 \gcc$, and $Y_e=.498$, after elapsed time of $2.3 \times 
        10^{-1}$ seconds ($\tnine=3.5$ and $\rho=2.0 \times 10^{6}\gcc$).  
        $Y_{net}$ are the network abundances, while $Y_{qse}$ are abundances 
        calculated assuming QSE from the network abundances of n, p, 
	\nuc{Si}{28}.}
    \label{fig:5c}
\end{figure}

As the expansion and cooling continues, this underabundance of light nuclei 
reverses.  By the time $2.9 \times 10^{-3}$ seconds have elapsed, with 
\tnine\ now 5.96 and $\rho=9.8 \times 10^{6} \gcc$, the abundances of the 
light group nuclei exceed their silicon quasi-equilibrium abundance by 
50\%.  Indeed, the abundances of most of the nuclei lighter than neon 
exceed their silicon quasi-equilibrium abundance, being nearly in 
quasi-equilibrium with the free nucleons.  Fig.  6b shows that the 
abundances of the light nuclei are also quite close to their NSE values, 
with free neutrons and \alp-particles slightly exceeding their NSE 
abundances.  This is the beginning of the \alp-rich freezeout, with the 
light group mass fraction of 38\% being 3\% larger than NSE predicts.  As a 
result, it is now the more massive members of the iron peak group (which 
comprises 50\% of the mass) whose abundance exceeds their NSE abundance.  
While the abundances of the nuclei between silicon and germanium display a 
relatively large 50\% spread when compared to NSE, the spread from 
quasi-equilibrium is only 5\%.  As the evolution proceeds the free nucleon 
and \alp-particle abundances continue to drop more slowly than NSE would 
predict.  By the time $\tnine= 5.0$ (elapsed time = $7.7 \times 10^{-2}$ 
seconds), the light group mass fraction has dropped to 13\%.  But for NSE 
at $\tnine=5.0$ and $\rho=5.8 \times 10^{6} \gcc$, the light group should 
contain less than 8\% of the mass.  As a result, \alp-particles have twice 
their NSE abundance, and the overabundance of the more massive members of 
the iron peak group grows.  However, as Figure 5b reveals, while the light 
QSE group has grown 4 orders of magnitude overabundant relative to QSE with 
silicon, the combined silicon and iron QSE groups remain within 5\% of 
mutual equilibrium.

\begin{figure}[t]
    \centering
    \includegraphics[angle=90,width=\textwidth]{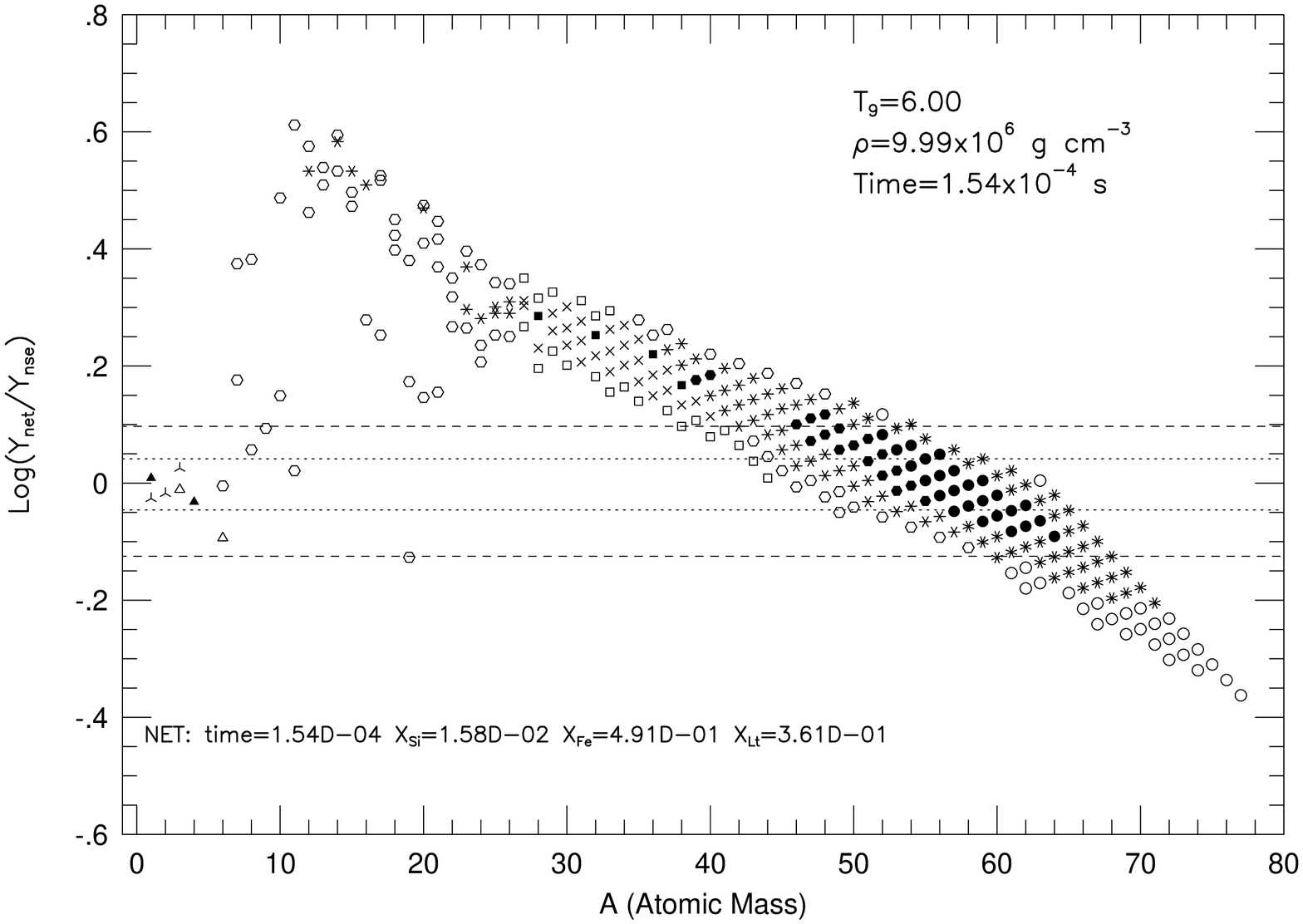}
    \caption{(a) Comparison of the network abundances to the NSE abundances 
        for a case with $\tninei=6$, $\rho_i=10^7 \gcc$, and $Y_e=.498$, after 
        elapsed times of $1.54 \times 10^{-4}$ seconds.  The filled shapes 
        represent network abundances larger than $10^{-6}$, and the hollow 
        shapes denote abundances less than $10^{-12}$.}
    \label{fig:6a}
    \addtocounter{figure}{-1}
\end{figure}

\begin{figure}[ht]
    \centering
    \includegraphics[angle=90,width=\textwidth]{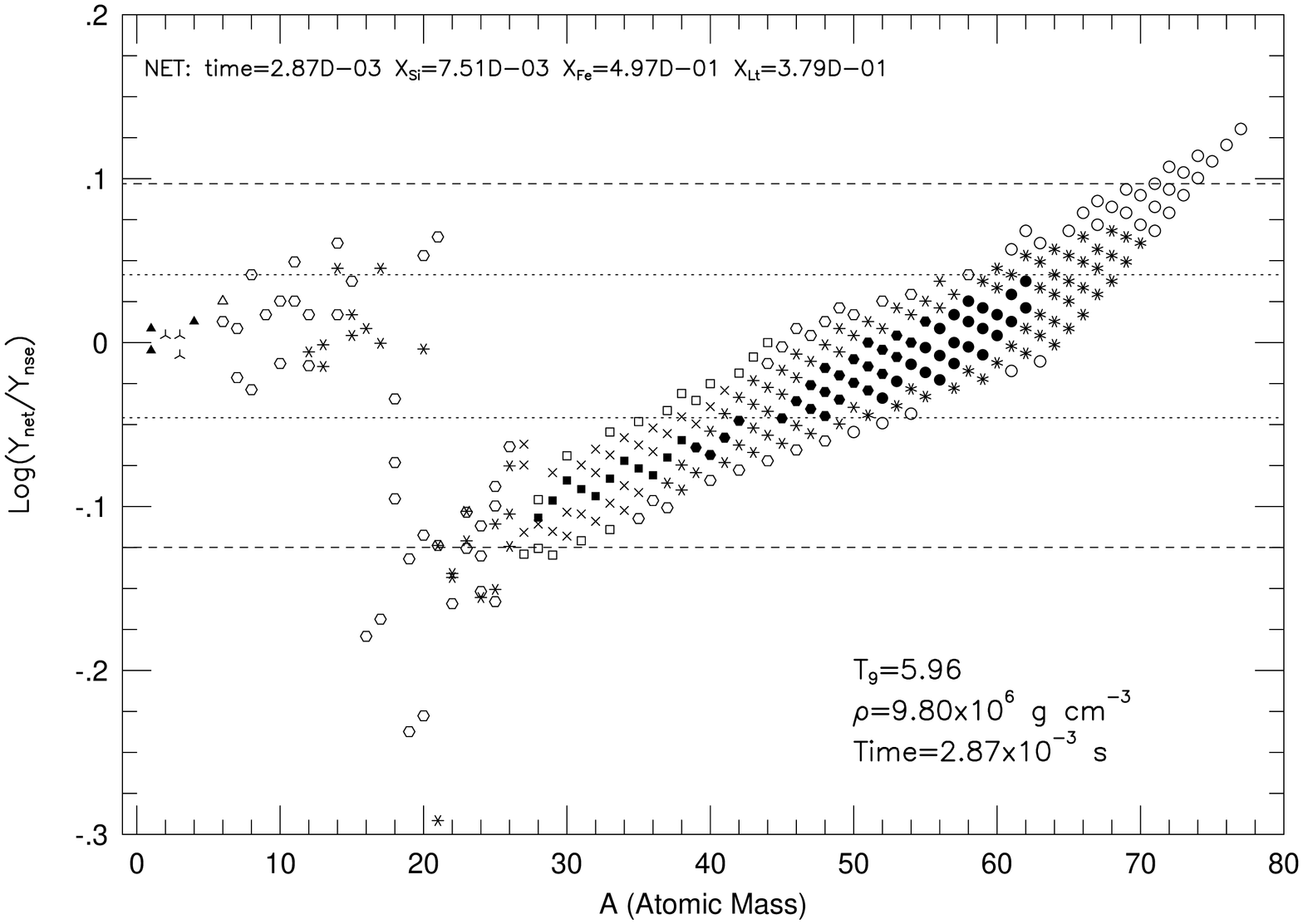}
    \caption{(b) Comparison of the network abundances to the NSE abundances 
	for a case with $\tninei=6$, $\rho_i=10^7 \gcc$, and $Y_e=.498$, after 
	elapsed time of $2.9 \times 10^{-3}$ seconds ($\tnine=5.96$ and 
	$\rho=9.8 \times 10^{6} \gcc$).  The filled shapes represent network 
	abundances larger than $10^{-6}$, and the hollow shapes denote 
	abundances less than $10^{-12}$.}
    \label{fig:6b}
    \addtocounter{figure}{-1}
\end{figure}

\begin{figure}[ht]
    \centering
    \includegraphics[angle=90,width=\textwidth]{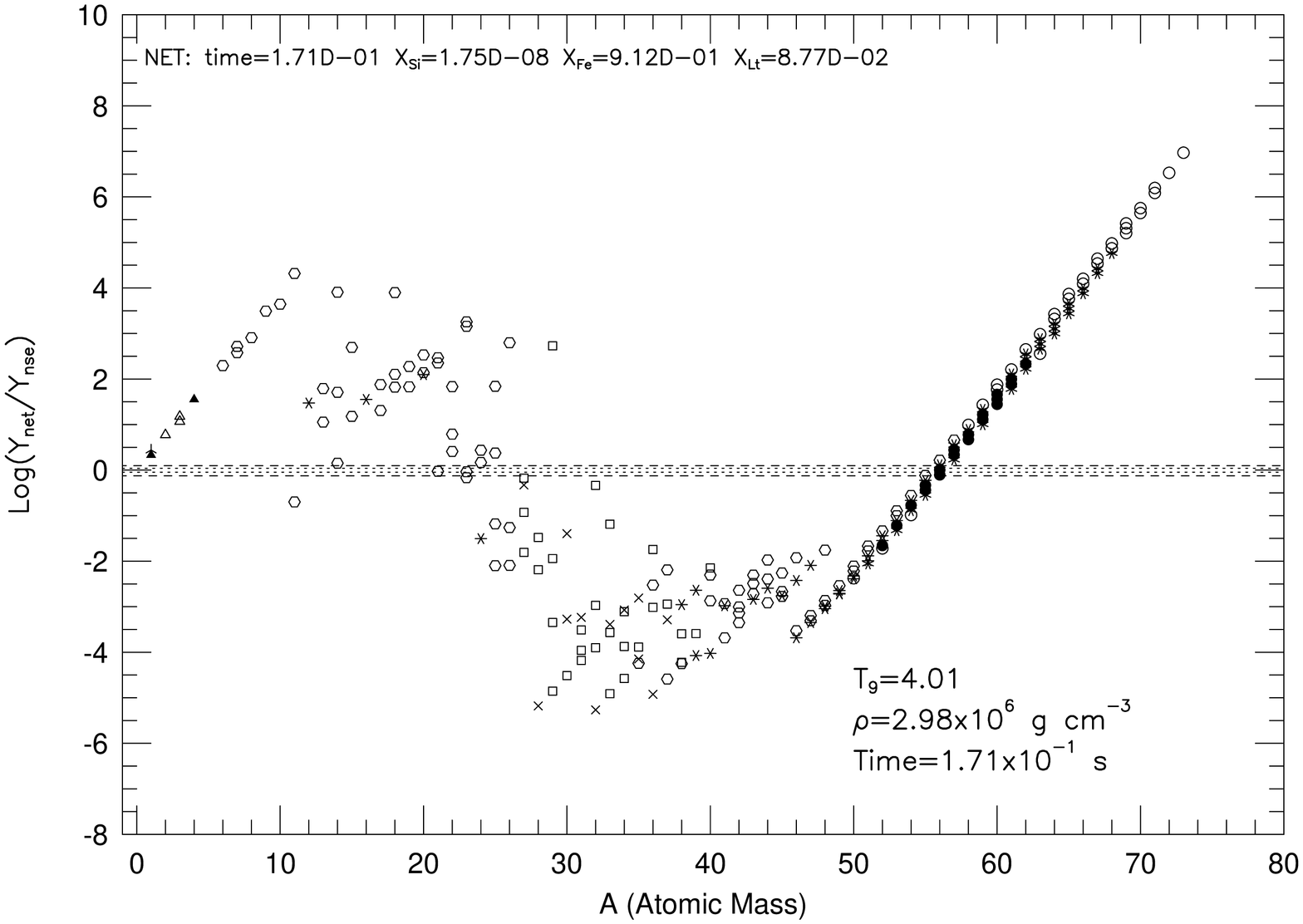}
    \caption{(c) Comparison of the network abundances to the NSE abundances 
	for a case with $\tninei=6$, $\rho_i=10^7 \gcc$, and $Y_e=.498$, after 
	elapsed time of $1.7 \times 10^{-1}$ seconds ($\tnine=4.0$ and 
	$\rho=3.00 \times 10^{6} \gcc$).  The filled shapes represent network 
	abundances larger than $10^{-6}$, and the hollow shapes denote 
	abundances less than $10^{-12}$.}
    \label{fig:6c}
\end{figure}

As time continues to elapse, and temperature continues to drop, the silicon 
group fragments into several smaller groups and detaches from the iron peak 
group.  The breakdown of this group results from the slow reaction flows 
from these nuclei which prevent the network abundances of the intermediate 
nuclei from declining as rapidly as quasi-equilibrium with the iron peak 
group would require.  With the total mass fraction of the former silicon 
group being a few $ \times 10^{-7}$, and the light and iron peak QSE groups 
containing 99.97\% of the mass, this breakdown into smaller 
quasi-equilibrium groups does not strongly affect the overall nuclear 
evolution.  The light and iron peak groups each remain unfragmented, with 
differences in relative abundance of less than 5\%.  With the light group 
mass fraction at $\tnine=4.0$ almost 17 times its NSE value and the 
\alp-particle abundance more than 35 times that predicted by NSE, the 
enrichment of the the more massive members of the iron peak group is 
enormous.  As Figure 6c shows, the abundances of the heaviest nuclei 
contained in our nuclear set, neutron-rich germanium nuclei, like 
\nuc{Ge}{73} are enhanced 10 million fold over NSE. Among more important 
nuclei, the abundances of the dominant isotopes of Cu and Zn are enhanced 
by 100 times, resulting in Cu and Zn mass fractions of .3\% and .1\%.  This 
behavior is largely consistent with that described by \cite{MeKC98}, who 
studied the related problem of the cooling of high entropy matter from NSE.

\begin{table}[t]
    \centering
    \small
    \caption{Selected abundances near freezeout for the \emph{\alp-rich 
    freezeout} example ($\tninei=6$, $\rho_1= 10^{7} \gcc$ and $Y_e=.498$)
    \label{tab:arf}}
    \begin{tabular}{c|ll|lll|l}
        \tableline
        Time &\multicolumn{2}{|c|}{2.28\ttt{-1}} &\multicolumn{3}{c|} 
        {2.93\ttt{-1}} &\multicolumn{1}{c}{4.45\ttt{-1}} \\
        \tnine &\multicolumn{2}{|c|}{3.50} &\multicolumn{3}{c|}{3.01} 
        &\multicolumn{1}{c}{2.09} \\
        $\rho$ &\multicolumn{2}{|c|}{1.99\ttt{6}} &\multicolumn{3} 
        {c|}{1.25\ttt{6}} &\multicolumn{1}{c}{4.25\ttt{5}} \\
        \tableline \tableline
        Nucleus 
        &\multicolumn{1}{c}{$Y_{net}$} &\multicolumn{1}{c|}{$Y_{qse}$}
        &\multicolumn{1}{c}{$Y_{net}$} &\multicolumn{1}{c}{$Y_{qse}$}
        &\multicolumn{1}{c|}{$Y_{nse}$} &\multicolumn{1}{c}{$Y_{net}$} \\
        \tableline
        \nuc{n }{ } & 1.94\ttt{-12}& 1.94\ttt{-12}& 5.11\ttt{-15}& 5.11\ttt{-15}
        & 1.91\ttt{-15} & 2.45\ttt{-17} \\
        \nuc{p }{ } & 2.60\ttt{-3} & 2.60\ttt{-3} & 7.15\ttt{-4} & 7.15\ttt{-4} 
        & 1.48\ttt{-5}  & 3.56\ttt{-6} \\
        \nuc{He}{4} & 2.05\ttt{-2} & 2.04\ttt{-2} & 2.05\ttt{-2} & 3.27\ttt{-2} 
        & 1.96\ttt{-6}  & 1.99\ttt{-2} \\
        \nuc{Si}{28}& 2.20\ttt{-10}& 2.20\ttt{-10}& 6.10\ttt{-10}& 6.10\ttt{-10}
        & 9.15\ttt{-8}  & 5.13\ttt{-9} \\
        \nuc{Si}{30}& 3.91\ttt{-12}& 6.44\ttt{-16}& 1.61\ttt{-11}& 2.95\ttt{-16}
        & 6.16\ttt{-15} & 2.81\ttt{-9} \\
        \nuc{S}{32} & 3.91\ttt{-10}& 2.31\ttt{-7} & 1.43\ttt{-9} & 3.80\ttt{-5} 
        & 3.41\ttt{-7}  & 3.29\ttt{-8} \\
        \nuc{S}{34} & 3.19\ttt{-13}& 1.71\ttt{-11}& 1.34\ttt{-12}& 7.96\ttt{-10}
        & 9.95\ttt{-13} & 2.03\ttt{-9} \\
        \nuc{Ar}{36}& 7.39\ttt{-10}& 8.80\ttt{-5} & 4.08\ttt{-9} & 7.16\ttt{-1} 
        & 3.84\ttt{-7}  & 2.83\ttt{-7} \\
        \nuc{Ar}{38}& 2.42\ttt{-13}& 4.18\ttt{-8} & 6.62\ttt{-13}& 1.32\ttt{-4} 
        & 9.91\ttt{-12} & 6.83\ttt{-10} \\
        \nuc{Ca}{40}& 7.57\ttt{-9} & 1.23\ttt{-1} & 2.24\ttt{-8} & 6.27\ttt{-4} 
        & 2.01\ttt{-6}  & 9.84\ttt{-7} \\
        \nuc{Ca}{42}& 3.60\ttt{-13}& 4.55\ttt{-6} & 5.12\ttt{-13}& 5.73\ttt{-1} 
        & 2.56\ttt{-12} & 2.86\ttt{-10} \\
        \nuc{Ti}{46}& 2.25\ttt{-12}& 6.72\ttt{-16}& 1.50\ttt{-12}& 8.47\ttt{-22}
        & 6.29\ttt{-10} & 1.34\ttt{-11} \\
        \nuc{Ti}{48}& 1.09\ttt{-17}& 2.00\ttt{-19}& 3.98\ttt{-19}& 9.31\ttt{-26}
        & 9.63\ttt{-15} & 2.72\ttt{-16} \\
        \nuc{Fe}{54}& 1.59\ttt{-5} & 1.60\ttt{-5} & 4.13\ttt{-7} & 4.14\ttt{-7} 
        & 1.10\ttt{-3}  & 2.38\ttt{-10} \\
        \nuc{Fe}{56}& 6.07\ttt{-9} & 6.10\ttt{-9} & 3.36\ttt{-11}& 5.33\ttt{-11}
        & 1.97\ttt{-8} & 3.31\ttt{-13} \\
        \nuc{Fe}{58}& 1.78\ttt{-16}& 1.79\ttt{-16}& 6.91\ttt{-20}& 1.17\ttt{-19}
        & 6.02\ttt{-18} & 2.22\ttt{-19} \\
        \nuc{Ni}{56}& 1.24\ttt{-2} & 1.24\ttt{-2} & 1.35\ttt{-2} & 1.35\ttt{-2} 
        & 1.54\ttt{-2}  & 1.36\ttt{-2} \\
        \nuc{Ni}{58}& 2.67\ttt{-3} & 2.68\ttt{-3} & 1.92\ttt{-3} & 3.07\ttt{-3} 
        & 4.86\ttt{-4}  & 1.43\ttt{-3} \\
        \nuc{Ni}{60}& 5.79\ttt{-7} & 5.82\ttt{-7} & 8.53\ttt{-8} & 2.22\ttt{-7} 
        & 4.89\ttt{-9}  & 6.67\ttt{-10} \\
        \nuc{Zn}{60}& 1.21\ttt{-5} & 1.21\ttt{-5} & 4.41\ttt{-5} & 7.04\ttt{-5} 
        & 4.77\ttt{-9}  & 3.67\ttt{-4} \\
        \nuc{Zn}{62}& 2.34\ttt{-5} & 2.35\ttt{-5} & 7.91\ttt{-5} & 2.06\ttt{-4} 
        & 1.94\ttt{-9}  & 2.34\ttt{-4} \\
        \nuc{Zn}{64}& 3.35\ttt{-8} & 3.37\ttt{-8} & 3.30\ttt{-8} & 1.40\ttt{-7} 
        & 1.84\ttt{-13} & 3.14\ttt{-9} \\
        \tableline
    \end{tabular}
\end{table}

In spite of the large concentration of light nuclei and large departure 
from NSE, as comparison of columns 2 and 3 of Table~\ref{tab:arf} 
demonstrates, the close adherence of the light and iron peak groups to 
quasi-equilibrium continues as the temperature drops below $\tnine=3.5$, on 
its way to freezeout.  This differs from the results of \cite{MeKC98} who 
claim that QSE breaks down below \tnine=4.  Though we agree that the single 
large QSE group has fragmented, QSE still applies to the groups which 
dominate the mass fraction and hence it is premature to say QSE has broken 
down.  As Figure 5c shows, these 2 tight QSE groups are linked across more 
than 30 orders of magnitude difference in quasi-equilibrium by a string of 
small grouplets with constant N and individual nuclei.  As in the previous 
cases, the continued decline of temperature toward \tnine=3 breaks QSE even 
within the light and iron peak groups.  As comparison of columns 4 and 5 of 
Table~\ref{tab:arf} shows, by an elapsed time of .293 seconds, the 
abundance of \nuc{He}{4} is 40\% smaller than QSE would predict based on
the network abundances of free protons and neutrons, with 
similar errors occurring among the most abundant nuclei in the iron peak 
group.  Unlike previous cases, while the free proton and neutron abundance 
decline markedly below \tnine=3.5, the large \alp\ mass fraction (8\%) 
remains virtually unchanged.  Within the time allowed by the cooling 
timescale, even the freezeout of the \alp\ producing photodisintegrations 
allows for a decline of $<5\%$ in the large \alp\ abundance.

Even this small relative change in the large abundance of \alp-particles 
has strong effects on the smaller abundances, like those of such important 
\alp-rich products as copper, zinc and germanium.  By the time the 
\alp-capture reactions which create these nuclei cease, represented in 
column 7 of Table~\ref{tab:arf} by abundances after an elapsed time of .445 
seconds (\tnine=2.1 and $\rho=4.3\ttt{5} \gcc$), the mass fractions of 
these species are .17\%, 3.8\% and .04\%, respectively.  Only those 
abundances which are large compared to the decline in $Y_{\alp}$ are 
unchanged by freezeout.  However, since it is these most abundant nuclei 
which dominate the energetics of nucleosynthesis, QSE (and future QSE based 
methods) can provide reasonable estimates of the energy release and other 
bulk nuclear properties necessary for accurate hydrodynamic simulations.  
In addition, though the QSE abundances frozen at \tnine=3 (column 5) 
underestimate the final abundances of the important \alp-rich iron peak 
nuclei by factors of a few to 10, these estimates are much more accurate 
than the corresponding NSE abundances (column 6).  Thus, even in the case 
of \alp-rich freezeout, where its predictions are the least reliable, 
quasi-equilibrium provides estimates of the nucleosynthesis which are 
adequate for accurate hydrodynamic modeling and much better than those 
provided by NSE.

\section{Conclusions}

Through these examples we have demonstrated that quasi-equilibrium 
continues to be a useful approximation during silicon burning even when the 
thermodynamic gradients are strong.  Provided there is sufficient abundance 
within a QSE groups, adjustments within the group happen much faster than 
reactions outside of these groups.  Furthermore these adjustments occur on 
the timescales of the unbalanced reactions, much more rapidly than 
reasonable thermodynamic variations.  In a recent paper (\cite{HKWT98}), 
we discuss methods which use QSE to reduce the 
computational load of silicon burning, while retaining the necessary 
accuracy, for the limited case of \alp-chain nucleosynthesis.  For 
relatively low temperatures, or fast expansions, conditions which result in 
the incomplete burning of silicon, quasi-equilibrium holds for temperatures 
exceeding approximately $3 \ttt{9}$ Kelvins.  Below this temperature, 
differential freezeout of reactions breaks up the large quasi-equilibrium 
groups but does not result in significant changes to the dominant 
abundances.  For conditions where temperature and density are sufficient to 
exhaust silicon, but the expansion occurs too rapidly for the complete 
incorporation of light nuclei into iron peak nuclei to occur, 
\emph{\alp-rich freezeout} occurs.  The resultant overabundance of 
\alp-particles and free nucleons obeys quasi-equilibrium, as does the 
enhancement of more massive iron peak nuclei.  Quasi-equilibrium fails in 
the vicinity of silicon once silicon is exhausted, but the light and iron 
peak groups which dominate the nuclear evolution continue to obey 
quasi-equilibrium until the temperature drops to roughly $3 \ttt{9}$ 
Kelvins.  Though the continued capture of the large abundances of light 
nuclei does result in significant abundance changes, predictions of 
abundance based on QSE provide good estimates of the largest abundances and 
hence the energetics and are significantly more reliable than those based 
on NSE. For expansion which occurs more slowly, previous authors have 
contended that silicon burned quickly to NSE, and that subsequently the 
nuclear distribution smoothly followed NSE until the NSE froze out at 
roughly $\tnine=3$.  We have found that while the abundances of the 
dominant nuclei, the iron peak nuclei, free nucleons and \alp-particles, do 
follow NSE rather smoothly, the intermediate nuclei, carbon, oxygen, and 
silicon, for example, lag behind.  As a result the abundances of these 
nuclei can be enhanced by several orders of magnitude at freezeout.  More 
importantly for our purposes, though NSE fails to properly account for 
these abundances, quasi-equilibrium remains an accurate estimate for these 
as well as the more important nuclei.  Thus quasi-equilibrium provides 
estimates of abundances as accurate or more accurate than NSE, even in the 
case of the \emph{normal freezeout}.  Taken together, we see that even for 
simulations which include strong thermodynamic gradients, quasi-equilibrium 
reliably provides good estimates of the abundances of many nuclear species 
resulting from silicon burning, and will be a valuable tool as we 
investigate the nucleosynthesis predictions of multi-dimensional supernovae 
simulations.

\acknowledgements

The authors would like to thank B.S. Meyer, C. Freiburghaus, K. Nomoto, 
J.C. Wheeler, and A.G.W. Cameron for fruitful discussions.  They would also 
like to thank the Institute of Theoretical Physics at the University of 
California, Santa Barbara, for its hospitality and support under NSF grant 
No.  PHY94-07194.  Work done at the Oak Ridge National Laboratory was 
supported by the U.S. Department of Energy under contract DE-FG02-96ER40983 
(Joint Institute for Heavy Ion Research) and DE-AC05-96OR22464 (with 
Lockheed Martin Energy Research Corp).  WRH was also supported in part 
by NASA Grant NAG5-2888.  FKT was supported in part by Swiss Nationalfonds 
grant 20-47252.96

\end{document}